%% file: main.tex
\documentclass[12pt]{iopart}

\pdfoutput=1
\expandafter\let\csname equation*\endcsname\relax
\expandafter\let\csname endequation*\endcsname\relax
\usepackage[english]{babel}
\usepackage{amsmath,
amssymb,amsfonts,latexsym}
\usepackage{graphicx}
\usepackage{mathrsfs} 
\usepackage{color}
\usepackage{booktabs}
\usepackage{cite}
\usepackage{bm}
\usepackage{braket}
\usepackage{nicefrac}
\usepackage{soul}
\usepackage{mathtools}
\usepackage{ulem}

\usepackage{tikz} 
\usepackage{pgfplots}
\usepgfplotslibrary{fillbetween}

\definecolor{Blue}{rgb}{0.00, 0.00, 1.00}
\definecolor{Red}{rgb}{1.00, 0.00, 0.00}
\definecolor{labelkey}{cmyk}{.1,.7,0.5,0}

\definecolor{mygreen}{rgb}{0,0.5,0}

\definecolor{blue(pigment)}{rgb}{0.2, 0.2, 0.6}
\usepackage{times}
\usepackage[colorlinks,
citecolor=blue,linkcolor=blue,urlcolor=blue]{hyperref}
\hypersetup{linktocpage}
\usepackage{scalerel}

\makeatletter
\def\@mkboth#1#2{}
\newlength\appendixwidth
\preto\appendix{\addtocontents{toc}{\protect\patchl@section}}
\newcommand{\patchl@section}{%
  \settowidth{\appendixwidth}{\textbf{Appendix }}%
  \addtolength{\appendixwidth}{1.5em}%
  \patchcmd{\l@section}{1.5em}{\appendixwidth}{}{\ddt}%
}
\makeatother

\def\eqref#1{(\ref{#1})}
\usepackage{cancel}

\newcommand{\bea}{\begin{eqnarray}}
\newcommand{\eea}{\end{eqnarray}}

\newcommand{\ord}[1]{\textbf{:}\, #1\, \textbf{:}}

\renewcommand*{\geq}{\geqslant}
\renewcommand*{\leq}{\leqslant}
\newcommand{\I}{\ensuremath{\mathbf{i}}}
\include{new_commands}

\graphicspath{{Figures/}}

\begin{document}

\title[]{Entanglement dynamics of a hard-core quantum gas during a Joule expansion}
\author{Filiberto Ares$^{1}$, Stefano Scopa$^{1}$ and Sascha Wald$^{2}$}
\address{$^1$ SISSA and INFN, via Bonomea 265, 34136 Trieste, Italy}
\address{$^2$ Statistical Physics Group, Centre for Fluid and Complex Systems, Coventry University, Coventry, England}
\date{\today}
\begin{abstract}
We study the entanglement dynamics of a one-dimensional hard-core quantum gas initially confined in a box of size $L$ with saturated density $\rho=1$. The gas is suddenly released into a region of size $2L$ by moving one of the box edges. We show that the analytic prediction for the entanglement entropy obtained from quantum fluctuating hydrodynamics holds quantitatively true even after several reflections of the gas against the box edges. We further investigate the long time limit $t/L\gg 1$ where a Floquet picture of the non-equilibrium dynamics emerges and hydrodynamics eventually breaks down.
\end{abstract}
\hrulefill
{\small \tableofcontents}
\hrulefill
\maketitle

\newpage
\section{Introduction}
The formulation of Generalized Hydrodynamics (GHD) (see e.g. Ref.~\cite{Bastianello2022} and Ref.~\cite{d-ls,Bertini2021,Alba2021,Bouchoule2021} for recent reviews) -- that is, an Euler hydrodynamic theory for one-dimensional integrable models -- is beyond doubt a recent milestone in the theory of out-of-equilibrium properties of many-body quantum systems. Its success is self-evident since GHD allows to easily obtain analytical results for the conserved charges and currents of interacting integrable models after a quantum quench, opening, in this way, a new possibility to move the research beyond the Tonks-Girardeau limit. It was originally formulated in Ref.~\cite{Bertini2016,Castro-Alvaredo2016} as an emergent large-scale theory of the front propagation dynamics obtained by joining together two Generalized Gibbs ensembles, extending previous results for the dynamics of such bi-partite states, see e.g. \cite{Antal1999,Schutz1999,Antal2008,Karevski2002,Platini2005,Platini2007,Collura2014a}. From this dawn, GHD has been extensively put to test against numerical simulations e.g. \cite{Piroli2017,Bulchandani2017,DeLuca2017,Doyon2017,Doyon2018,Bulchandani2018,Bastianello2019,Bouchoule2020a,Moller2020}, cold atom experiments \cite{Schemmer2019,Malvania2020} and by relaxing some hypothesis that could have undermined its validity (see e.g. Ref.~\cite{Bastianello2020a,Bastianello2021,DeNardis2018,DeNardis2019,Bouchoule2020b}), but nonetheless it always returned a predictive power beyond expectations. On the other hand, it is also known that GHD is a semi-classical theory and so it can only act as background for some quantum fluctuations in the analysis of intrinsically quantum features, such as entanglement. However, a reliable description of the system is obtained by focusing on its low-energy regime and by building a conformal field theory on top of the GHD for the description of the missing large-scale quantum fluctuations. This method is generically referred to as quantum fluctuating hydrodynamics (or, more simply, as quantum GHD) see Ref.~\cite{Ruggiero2019,Ruggiero2020,Scopa2021a,Scopa2021b,Ruggiero2022,Rottoli2022}.\\

As a consequence, some simple platforms for the study of quantum transport with non-interacting spin chains (or, equivalently, with impenetrable quantum particles) gained a renewed interest, especially those in which a scaling behavior of the propagating fronts was observed long ago, see e.g. Ref.~\cite{Kagan1996,Minguzzi2005,Leach1982,Antal1999,Schutz1999,Antal2008,Karevski2002,Rigol2004,Platini2005,Platini2007,Rigol2015}. In fact, if such a scaling theory turns out to be essentially equivalent to a non-interacting limit of GHD, a careful usage of quantum fluctuating hydrodynamics is instead needed to make quantitative predictions on the entanglement dynamics beyond numerical and heuristic considerations \cite{Eisler2008,Vicari2012,Alba2014,Gruber2019}.\\

In this context, we shall investigate the entanglement dynamics after a geometric quench where the boundary region  of a trapped one-dimensional gas is suddenly changed \cite{Caux2012,Collura2013,Collura2013b,Dubessy2020}. In particular, we wish to understand whether the presence of boundaries for the expanding gas will eventually spoil the validity of the hydrodynamic approach and, if so, when this is expected to happen. An ideal testbed for these ideas is the so-called Joule expansion \cite{Collura2013,Collura2013b,Camalet2008,Joule-exp1,Joule-exp2,Grochowski2020,Grochowski2021}, where the system undergoes a sudden expansion of the containing tank as detailed in the following section. For this problem, we characterize the entanglement dynamics across the reflections in the Euler scaling limit and we present a Floquet picture for its long time limit. We also show that this quench protocol gives the same set of conserved quantities of an infinite system prepared in a periodic fashion and let to relax freely. But this analogy does not extend to the entanglement properties. We further discuss the failure of hydrodynamics at times much larger than the system size and the emergence of revivals of the initial configuration.\\

\noindent
{\it Oultine.} In Sec.~\ref{sec:model}, we introduce the model and the quench protocol considered in this work. Afterwards, in Sec.~\ref{sec:phase-space-dyn}, we construct a large-scale description of the model in terms of the local occupation function of fermionic modes and we study its time evolution in phase space. With this method, we reconstruct the behavior of the conserved quantities during the quench dynamics. In Sec.~\ref{sec:quantum-hydro}, we proceed with the re-quantization of the theory and with the calculation of the entanglement entropy, following the discussion of the recent Ref.~\cite{Scopa2021a,Scopa2021b,Ruggiero2022}. Finally, in Sec.~\ref{sec:long-time}, we discuss the long time limit and the breakdown of hydrodynamics. In Sec.~\ref{sec:conclusion}, we report our conclusions while details on the calculation of the Fermi contour and on the numerical implementation are left to \ref{app:Fermi-contour} and \ref{app:numerics} respectively.

\section{The model and the quench protocol}\label{sec:model}
We consider a one-dimensional gas of hard-core quantum particles with nearest neighbor 
hopping and coupled to a potential $V$. This system is described by the Hamiltonian
\begin{align}\label{eq:model}
\Ha=-\frac{1}{2}\sum_{j\in\mathbb{Z}} \left(\hat{c}^\dagger_{j}\hat{c}_{j+1} +\hat{c}^\dagger_{j+1}\hat{c}_j\right)+ V(j,t) \hat{c}^\dagger_j \hat{c}_j .
\end{align}
Here, $\hat{c}^\dagger_j$ and $\hat{c}_j$ denote the creation and annihilation operator of 
a spinless fermion at lattice site $j$, satisfying the standard
canonical anticommutation relations $\{ \hat{c}_i,\hat{c}^\dagger_j\}=\delta_{ij}$.
 We further consider the situation in which the model is initially prepared in the ground state of the Hamiltonian \eqref{eq:model} with a box potential $V={\cal V}_{[a,b]}$ of infinitely high edges
\be
V(j,t< 0)= {\cal V}_{[-L,0]}(j)\equiv\lim_{\Lambda\to\infty}\begin{cases} -1, \quad \text{if $j\in[-L,0]$};\\[4pt]
\Lambda, \quad\text{otherwise} 
\end{cases}
\ee
\begin{figure}[t]
\centering
\includegraphics[width=.5\textwidth]{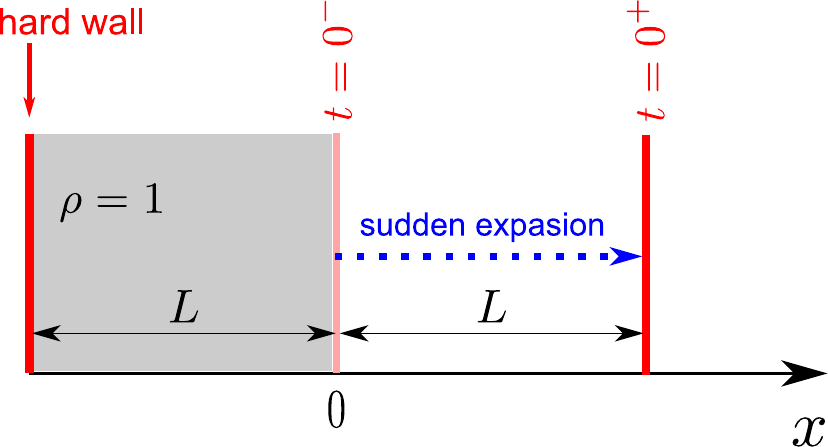}
\caption{Illustration of the quantum Joule expansion quench protocol considered in this work: We initially load a hard-core quantum gas in the region $-L\leq j\leq 0$ with density $\rho=1$. At time $t=0^+$ we suddenly move the right edge of the box confinement from position $j=0$ to position $j=L$ and  we study the unitary dynamics in the doubled box generated by the hopping fermionic Hamiltonian in Eq.~\eqref{eq:model}.}\label{fig:protocol}
\end{figure}
which is the product state
\be\label{eq:initialWF}
\ket{\Psi(0)}=\left(\bigotimes_{j<-L} \ket{0}_j  \right) \otimes \left(\bigotimes_{j\in [-L,0]} \ket{1}_j  \right) \otimes\left(\bigotimes_{j\in >0} \ket{0}_j \right),
\ee
where $\ket{n=0,1}_j$ are the eigenstates of the number operator $\hat{c}^\dagger_j\hat{c}_j$ with eigenvalues $0,1$ respectively. 

At time $t=0^+$ we suddenly double the size of the initial box confinement, i.e. $V(j,t>0)={\cal V}_{[-L,L]}$, and we investigate the unitary dynamics $\ket{\Psi(t)}=\exp(-\I t \Ha)\ket{\Psi(0)}$ generated by the Hamiltonian \eqref{eq:model} in the expanded region of size $2L$, see Fig.~\ref{fig:protocol} for an illustration.
We shall refer to this quench protocol as {\it quantum Joule expansion} (QJE), 
in analogy to the well-known classical process in statistical mechanics.
Intuitively, the non-equilibrium dynamics during the QJE can be divided in different regimes.
First, for times $t\leq L$, we observe the free expansion of a hard-core quantum gas 
towards the right vacuum. This free expansion
has been thoroughly investigated in the literature, see e.g. 
Refs.~\cite{Antal1999,Antal2008,Karevski2002,Platini2005,Platini2007} 
and Refs.~\cite{Eisler2008,Vicari2012,Alba2014,Eisler2016,Eisler2018,Gruber2019,Eisler2020,Eisler2021,Gruber2021,Dubail2017,Scopa2021a}
for studies on the entanglement. 
As we shall see further below, at time $t=L$ the first front of excitations 
arrives at the infinite potential well and is reflected. Thus, for $t>L$
we observe a regime where the dynamics of the quantum gas is affected by
multiple reflections against the left and right walls located at the positions $j=\pm L$.
While the non-equilibrium profiles of charges and currents after such reflections 
has been understood in literature using semi-classical hydrodynamic pictures
(see Sec.~\ref{sec:phase-space-dyn} below and, e.g., Ref.~\cite{Platini2007}), we 
 investigate the behavior of intrisically quantum properties such as entanglement. To this end, in Sec.~\ref{sec:quantum-hydro}, we shall employ the recently 
developed framework of quantum fluctuating hydrodynamics for the exact asymptotic 
calculation of the R\'enyi entropies $S_n(j,t)$ of a bipartition $A\cup B$, $A=[-L,j]$ 
and $B=[j+1,L]$, defined as
\be\label{eq:Renyi-def}
S_n(j,t)=\frac{1}{1-n}\log\tr\hat\rho_A(t)^n, \quad n \in\mathbb{N},
\ee
where $\hat\rho_A(t)\equiv\tr_B\ket{\Psi(t)}\bra{\Psi(t)}$ is the reduced density 
matrix of subsystem $A$ at a certain time $t$. From Eq.~\eqref{eq:Renyi-def}, the von 
Neumann entanglement entropy $S_1(j,t)=-\tr \hat\rho_A(t)\log\hat\rho_A(t)$ is obtained 
by taking the analytic continuation of $S_n(j,t)$ for $n\to 1$.

\section{Hydrodynamics in the phase space}\label{sec:phase-space-dyn}
Although the non-interacting nature of the quench problem under analysis might call for an
exact lattice description, we rather consider the non-equilibrium dynamics of the QJE
in the Euler scaling regime $j,t\to \infty$ at fixed $j/t = \text{cst}$.
In this regime, a hydrodynamic description of the initial state in terms of the 
local Fermi occupation number $n(x,k)$ yields asymptotically exact results 
for the conserved charges and currents of the model. In this formulation,
the initial state~\eqref{eq:initialWF} corresponds to a macrostate 
that can be expressed by the local Fermi occupation function~\cite{Karevski2002,Antal2008,Wendenbaum2013,Scopa2021a}
\be\label{eq:initial-macrostate}
n_0(x,k)=\begin{dcases} 1, \quad\text{if $x\in[-L,0]$ and $-\pi\leq k\leq \pi$,}\\[4pt]
0, \quad\text{otherwise},
\end{dcases}
\ee
where we replaced the lattice position $j\in\mathbb{Z}$ with a continuum spatial 
variable $x\in\mathbb{R}$.
Hence, the system is decomposed into 
non-interacting fermionic modes that entirely fill the region $-L\leq x \leq 0$, 
and leave the rest of the system empty.
As  the particles populating 
the initial macrostate in Eq.~\eqref{eq:initial-macrostate} are non-interacting,
one can determine the time evolution of $n_0(x,k)$ by simply following the free
evolution $x_t=x+v(k)t$ of each particle with momentum $k$ from its initial position $x$.
This yields
\be\label{eq:n_t}
n_t(x,k)=n_0(x-v(k) t, k),
\ee
with velocity $v(k)=\sin k$ for the free Fermi gas~\cite{Fagotti2017,Fagotti2020}.

Equivalently, due to the zero-entropy condition $n_0=\{0,1\}$ of Eq.~\eqref{eq:initial-macrostate} which is preserved by the time evolution (cf Eq.~\eqref{eq:n_t}), one can conveniently encode the information on the phase space dynamics in a local (split-)Fermi sea $\Gamma_t(x)=\bigcup_{j=1}^Q[k_{F,2j-1},k_{F,2j}]$, with local Fermi points $k_{F,j}(x,t)$ given as solution of the zero-entropy hydrodynamic equation \cite{Doyon2017}
\be\label{eq:GHD}
\left(\de_t + \sin k_{F,j} \de_x\right) k_{F,j}=0, \qquad j=1,\dots,2Q.
\ee
In terms of $\Gamma_t(x)$, one can express the fermionic occupation function \eqref{eq:n_t} as
\be
n_t(x,k)=\begin{cases} 1, \quad \text{if  $k\in \Gamma_t(x)$};\\[4pt]
0, \quad \text{otherwise}\end{cases}
\ee
and determine the profile of a conserved quantity $q$ as
\be\label{eq:charges}
q(x,t)=\int_{-\pi}^\pi \frac{\dd k}{2\pi} n_t(x,k) \ h_q(k) = \int_{\Gamma_t(x)} \frac{\dd k}{2\pi} \ h_q(k),
\ee
where $h_q(k)$ is the single-particle eigenvalue associated to $q$. For concreteness, one finds that $h_e= -\cos k$ for the energy density $e$, $h_{J_\rho}(k)=\sin k$ for the particle current $J_\rho$, $h_\rho= 1$ for the particle density $\rho$ and similarly for other quantities.\\

We first consider the regime $t\leq L$, characterized by the free expansion of the quantum gas without reflections of the Fermi points. Since the propagating modes on the initial Fermi contour are only those belonging to the position $x=0$, it is easy to see that the entanglement spreads during the quench dynamics inside a region $|x|\leq t$, the so-called {\it light cone}, which is determined by the fastest particles $k=\pm\pi/2$ with velocity $v=\pm 1$. Hence, at position $0\leq x \leq t$, it is easy to show that Eq.~\eqref{eq:GHD} is characterized by the two solutions 
\be\label{eq:gamma-0refl}
\Gamma_t(x)\equiv \left[\arcsin\frac{x}{t}; \pi-\arcsin\frac{x}{t}\right]
\ee
while outside the light-cone
\be\label{eq:gamma-prop2}
\Gamma_t(x)=\emptyset, \quad \text{for} \ t>x\geq 0.
\ee
An analogous treatment applies for positions $-L \leq x \leq 0$ by exploiting the particle-hole symmetry of the problem. This leads to the useful identity
\be\label{eq:gamma-prop1}
\Gamma_t(-x)=[-\pi,\pi] \setminus \Gamma_t(x), \quad \forall \  x\in[-L,L] \, .
\ee

Using Eq.~\eqref{eq:gamma-0refl}-\eqref{eq:gamma-prop2} in Eq.~\eqref{eq:charges} with $h_\rho\equiv 1$, one obtains the asymptotic density profile of the expanding gas before reflections as \cite{Antal1999,Antal2008,Allegra2016,Dubail2017,Scopa2021b}
\be\label{eq:dens-0refl}
\rho(x,t\leq L)=\begin{cases}\bar\rho(x,t)\equiv\arccos(x/t)/\pi; \quad -t\leq x \leq t;\\[4pt]
1, \quad -L\leq x<-t;\\[4pt]
0, \quad \text{otherwise.}\end{cases}
\ee

\begin{figure}[t]
\centering
\includegraphics[width=0.75\textwidth]{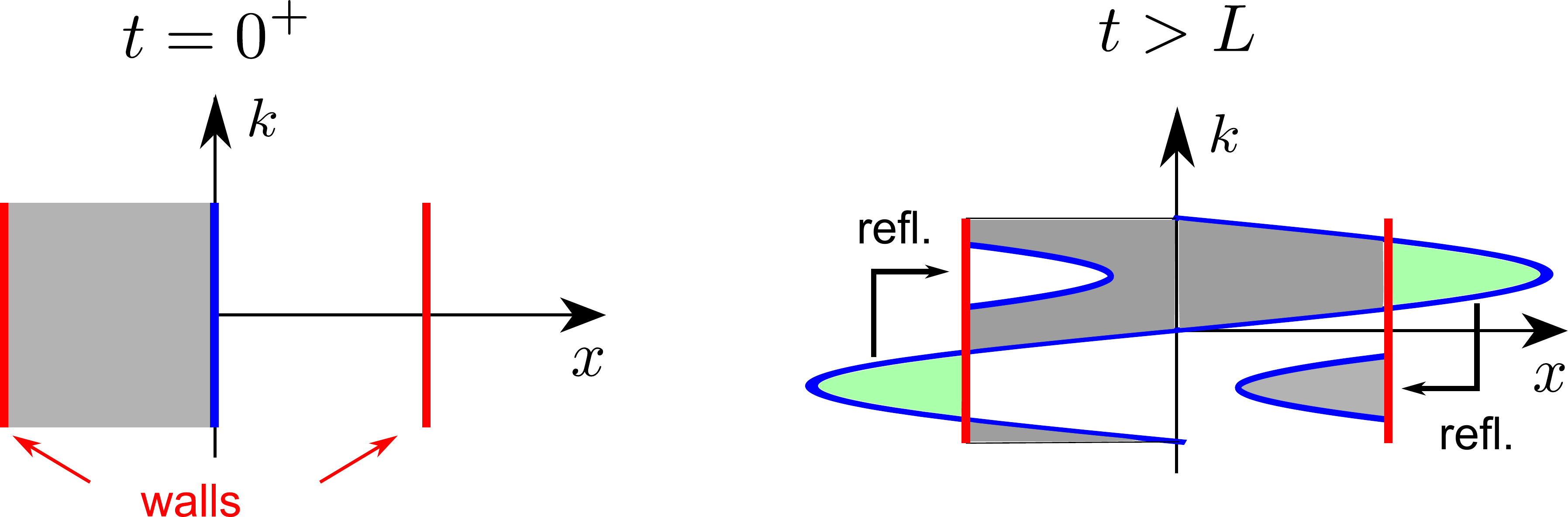}
\caption{Local occupation number $n_t(x,k)$ of the free Fermi gas in the plane $x$-$k$ ({\it gray region}) and the associated Fermi contour $\Gamma_t$ ({\it thick blue line}) at time $t=0$ (left panel) and after the first reflection $t>L$ (right panel). The green regions mark the reflected modes $k\to -k$ in the hydrodynamic picture.}\label{fig:Gamma-reflections}
\end{figure}

At times $t>L$ , the system will undergo multiple reflections at the left and right walls $x=\pm L$. Due to the absence of interactions and to the particularly simple choice of potential, each particle with momentum $k$ is elastically reflected at the wall and it simply inverts its trajectory, $k\to -k$. This leads to the phase space picture depicted in Fig.~\ref{fig:Gamma-reflections}.\\
With this picture at hand, one can analytically determine the structure of the Fermi contour $\Gamma_t(x)$ for given position $x$ and time $t$. Focusing on $x>0$, one can write
\be\label{eq:Gamma-refl}
\Gamma_t(x)=\bigcup_{j=1}^{N_++N_-} [k_{F,2j-1},k_{F,2j}]
\ee
with
\be
N_-=\left\lfloor{\frac{t+x}{2L}}\right\rfloor, \quad N_+=\left\lceil{\frac{t-x}{2L}}\right\rceil
\ee
where $\lfloor x\rfloor$ (resp.~$\lceil x \rceil$) is the standard floor (resp.~ceil) rounding of $x$ to integer value, and Fermi points defined as
\be\label{eq:Fermi-points-refl}
k_{F,\nu}(x,t)=\begin{cases} 
\arcsin\frac{x+2L(\nu-1)}{t}, \quad \nu=1,\dots, N_+;\\[4pt]
\pi-\arcsin\frac{x+2L(\nu-N_+-1)}{t}, \quad\nu=N_++1,\dots,2N_+; \\[4pt]
\arcsin\frac{x+2L(2N_+-\nu)}{t}, \quad \nu=2N_++1,\dots, 2N_+ + N_-;\\[4pt]
-\pi -\arcsin\frac{x-2L (2Q-\nu+1)}{t}, \qquad \nu=2 N_+ +N_- +1, \dots,2Q,\\[4pt]
\end{cases}
\ee
sorted in increasing order to form the set $\{k_{F,j}(x,t)\}_{j=1}^{2Q}$, with total number of Fermi points $2Q=2(N_+ + N_-)$.\\

 For $x<0$, the Fermi contour can be obtained from Eq.~\eqref{eq:Gamma-refl} using \eqref{eq:gamma-prop1}. With our notations, it is easy to recover the result of Eq.~\eqref{eq:gamma-0refl} since for $t\leq L$ one finds that $N_-=0$ and $N_+=1$.
Eqs.~\eqref{eq:Gamma-refl}-\eqref{eq:Fermi-points-refl} are obtained with simple but lengthy algebra that we report in~\ref{app:Fermi-contour}.\\

\begin{figure}[t]
\centering
\includegraphics[width=\textwidth]{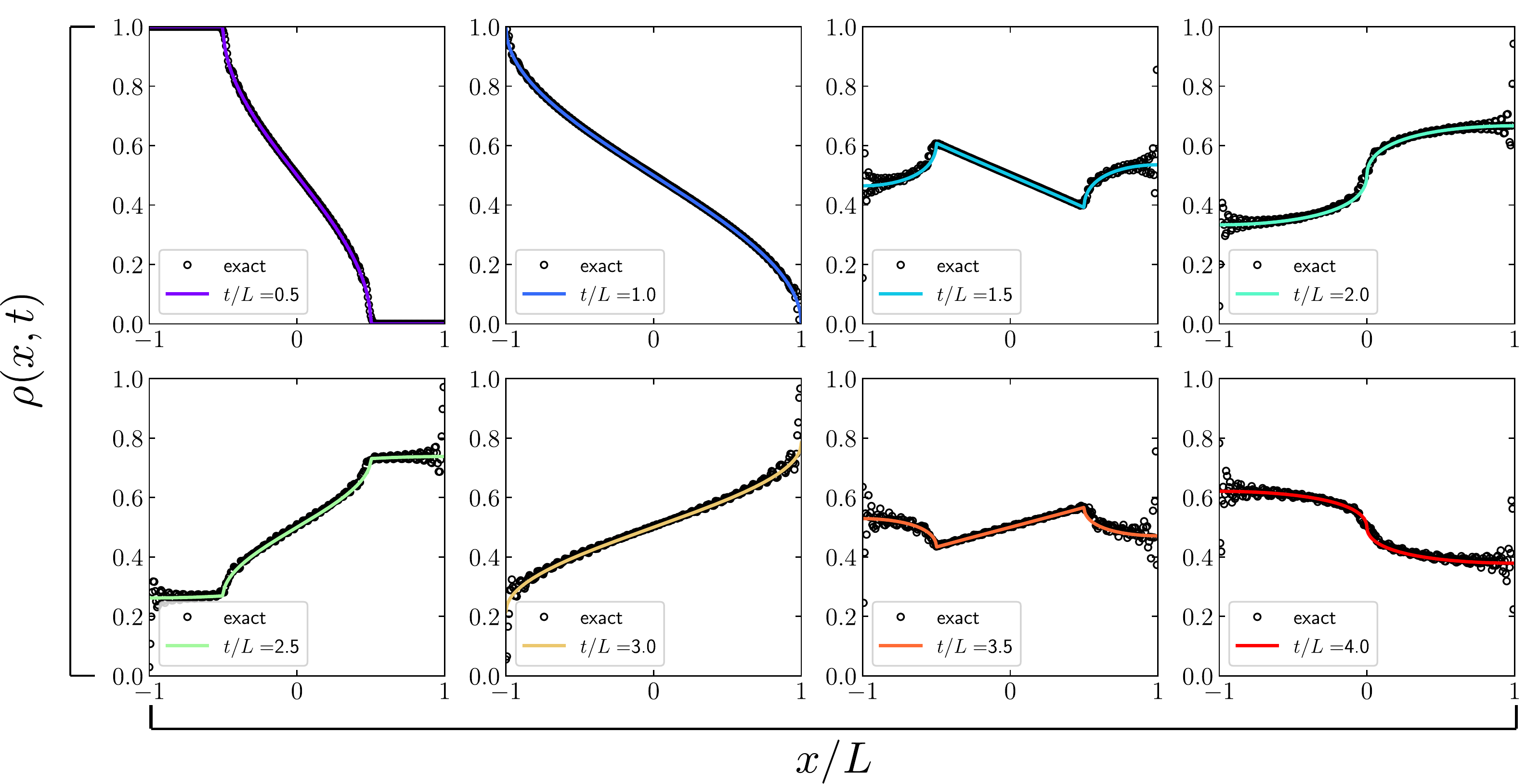}
\caption{Particle density profile during a quantum Joule expansion at different times (panels) as a function of the rescaled position $x/L$. The asymptotic prediction ({\it solid line}) in Eq.~\eqref{eq:dens-refl} is compared with exact numerical calculations ({\it symbols}) performed on a lattice with $300$ sites. The agreement in each panel is extremely good.}\label{fig:test-density}
\end{figure}

From the knowledge of $\Gamma_t(x)$, we evaluate the particle density using Eq.~\eqref{eq:charges} as 
\be\label{eq:dens-refl}
\rho(x,t)=\sum_{j=1}^{Q} \frac{k_{F,2j-1}-k_{F,2j}}{2\pi}.
\ee
In Fig.~\ref{fig:test-density}, we test our result against exact numerical calculations for the lattice model finding an excellent agreement. We briefly discuss the implementation used for the numerical calculations in \ref{app:numerics}.

\subsection{Analogy with the multiple domain wall setting}
\begin{figure}[t]
\centering
\includegraphics[width=0.75\textwidth]{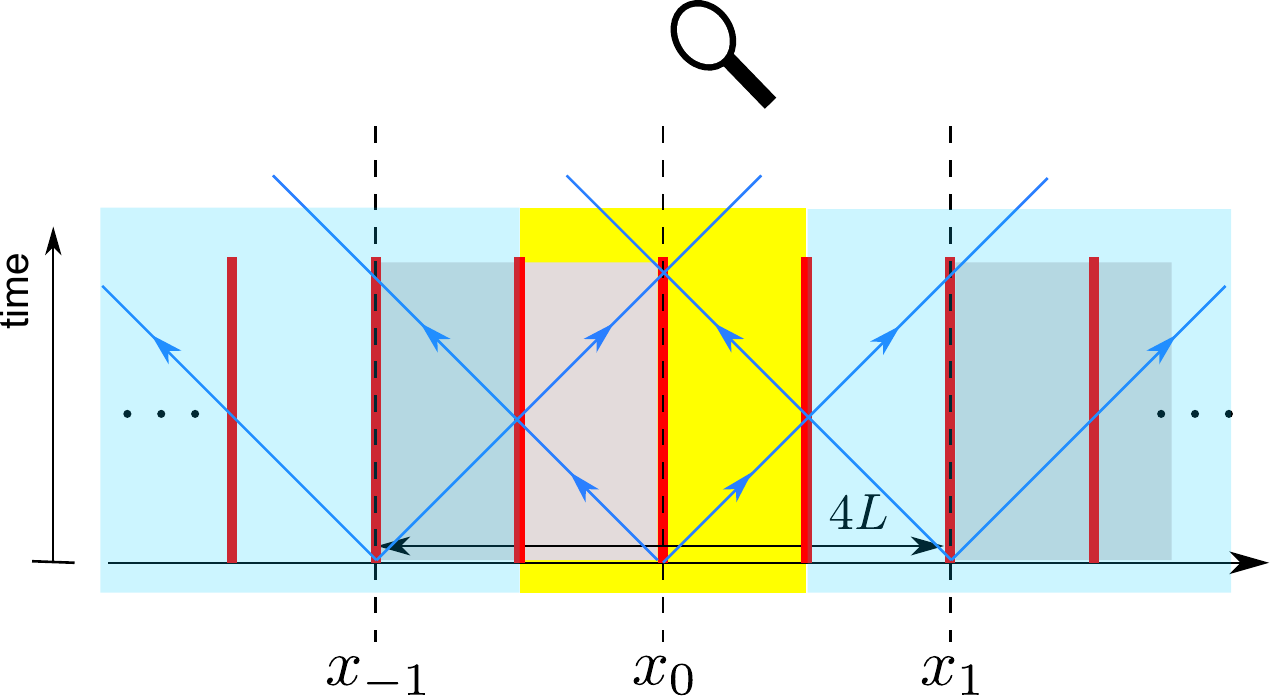}
\caption{Analogy of the phase space picture between the QJE and the MDW configuration. By focusing on the region $|x|\leq L$ ({\it yellow area}), one can equivalently interpret the superposition of the light cones (depicted with solid lines) as result of reflections against walls at $x=\pm L$ (QJE) or as the propagation of particles coming from other junctions located at $x_p=2pL$, $p\in\mathbb{Z}$ in absence of confinement (MDW).}\label{fig:multipleDW}
\end{figure}

It is instructive to notice that the phase space picture in Fig.~\ref{fig:Gamma-reflections} obtained for the QJE at times $t>L$ can be equivalently interpreted as the result of an initial configuration
\be\label{eq:nMDW}
n^{(\text{copy})}(x,k)\equiv \tilde{n}(x \ {\rm mod} \ 4L, k), \; \tilde{n}(|x|\leq 2L ,k)=\begin{cases} 1, \qquad\text{if $|k|\leq \pi$ and $x\in [-2L,0]$};\\[4pt]
0, \; \text{otherwise}\end{cases}
\ee
where the initial domain wall configuration of Sec.~\ref{sec:phase-space-dyn} is doubled and copied over an infinite one-dimensional lattice with $4L$ periodicity in absence of any confinement for $t\geq 0^+$, see Fig.~\ref{fig:multipleDW} for an illustration. We will often refer to this configuration as {\it multiple domain wall setting} (MDW).\\

From this analogy, the density profile in Eq.~\eqref{eq:dens-refl} can be also written as the sum of the density profiles coming from the junctions at $x_p=2pL$, $p\in\mathbb{Z}$ and propagating up to position $|x|\leq L$ at time $t$ in absence of reflections (cf Eq.~\eqref{eq:dens-0refl}) \cite{Collura2013,Collura2013b}
\be\label{eq:dens-copy}
\rho(x,t)=\bar\rho(x,t) - \sum_{p=1}^{N_-}(-1)^{p} \bar\rho(x_p-x,t)+ \sum_{p=1}^{N_+} (-1)^p\bar\rho(x_p+x,t),
\ee
where we recall $\bar\rho(x,t)=\arccos(x/t)/\pi$ and $N_\mp=\lfloor \frac{ t\pm |x|}{2L} \rfloor$ is the farther particles source that contribute to the charges profile at given $x$ and $t$, see~\ref{app:Fermi-contour}.
\section{Entanglement dynamics}\label{sec:quantum-hydro}
In this section, we wish to investigate the entanglement spreading during the quantum Joule expansion. We notice that for the non-homogeneous quench problem under analysis, standard lattice and field theory techniques are typically not valid and, although being highly desirable, a quantum hydrodynamic description obtained directly from the lattice model in Eq.~\eqref{eq:model} is still out-of-reach. However, a fruitful possibility is established by the recently developed framework of quantum fluctuating hydrodynamics \cite{Ruggiero2019,Ruggiero2020,Scopa2021a,Scopa2021b,Rottoli2022,Ruggiero2022} where relevant quantum processes in the low-energy regime are modeled as  a Luttinger liquid on top of the phase space hydrodynamics that we previously determined. More precisely, it is expected that relevant quantum processes at low energy are in the form of particle-hole excitations around the Fermi points and can be conveniently described introducing a density fluctuating field
\be
\delta\hat\rho=\frac{1}{2\pi} \de_x \hat\phi,
\ee
which is typically further expressed as the derivative of a bosonic field $\hat\phi$. Following the standard bosonization (see e.g. Ref.~\cite{Giamarchi2007,Cazalilla2004,Cazalilla2011}), we expand the lattice fermionic operator in terms of vertex operators of the bosonic field obtaining
\be
\hat{c}_x^\dagger(t)\propto \ord{\exp\left(\frac{\I}{2}(\hat\phi_+(x,t)-\hat\phi_-(x,t))\right)}+\text{subleading terms}
\ee
up to a semi-classical phase that is unimportant for our scopes (see e.g. Ref.~\cite{Allegra2016,Dubail2017,Ruggiero2019} for more details) and at leading order in the scaling dimension of the field theory operators. Here, we considered a chiral decomposition of the bosonic field $\hat\phi=\hat\phi_+ + \hat\phi_-$ where $\hat\phi_-$ (resp.~$\hat\phi_+$) is the left- (resp. right) moving component of the fluctuating field along the Fermi contour. Crucially, the effective Hamiltonian that governs the dynamics of the quantum fluctuations is that of a non-homogeneous Luttinger liquid \cite{Dubail2017,Brun2017,Brun2018,Ruggiero2019,Bastianello2020,Scopa2020,Scopa2021a,Scopa2021b,Rottoli2022,Ruggiero2022}
\be\label{eq:H-LL}
\Ha_{LL}[\Gamma]=\frac{1}{2\pi} \int_{\Gamma} \frac{\dd \theta}{2\pi} \ {\cal J}(\theta) \sin k(\theta) \left(\de_\theta\hat\phi_{a(\theta)}(\theta)\right)^2, \qquad (a(\theta)\equiv \pm={\rm sign}(k(\theta)))
\ee
here reported in terms of a coordinate $\theta$ that parametrizes the Fermi contour $\Gamma$ as a simple curve in the $x$-$k$ plane, ${\cal J}(\theta)$ is the Jacobian factor associated to the change of coordinates $x\mapsto \theta$, see also Ref.~\cite{Scopa2021a} for more details.

We now apply this framework in the analysis of entanglement dynamics for the specific quench setting discussed in Sec.~\ref{sec:model}. In particular, due to the non-interacting nature of the problem considered, quantum fluctuations in the initial state are obtained as the ground state of the Luttinger liquid Hamiltonian \eqref{eq:H-LL} with Fermi contour 
\be\label{eq:initial-Gamma}
\Gamma_0=\{ (x,k) \ : \ x=0;\;  -\pi\leq k \leq \pi\},
\ee
which allows for a very intuitive parametrization as $\theta\equiv k$. Furthermore, due to the non-interacting nature of the underlying quench protocol, the effect of the time evolution is simply to transport these quantum fluctuations along the curve $\Gamma_t$ that gets deformed in time according to the semi-classical hydrodynamics of Sec.~\ref{sec:phase-space-dyn}.

\subsection{Quantum fluctuations and entanglement spreading}\label{sec:entanglement-calc}
\begin{figure}[t]
\centering
\includegraphics[width=0.8\textwidth]{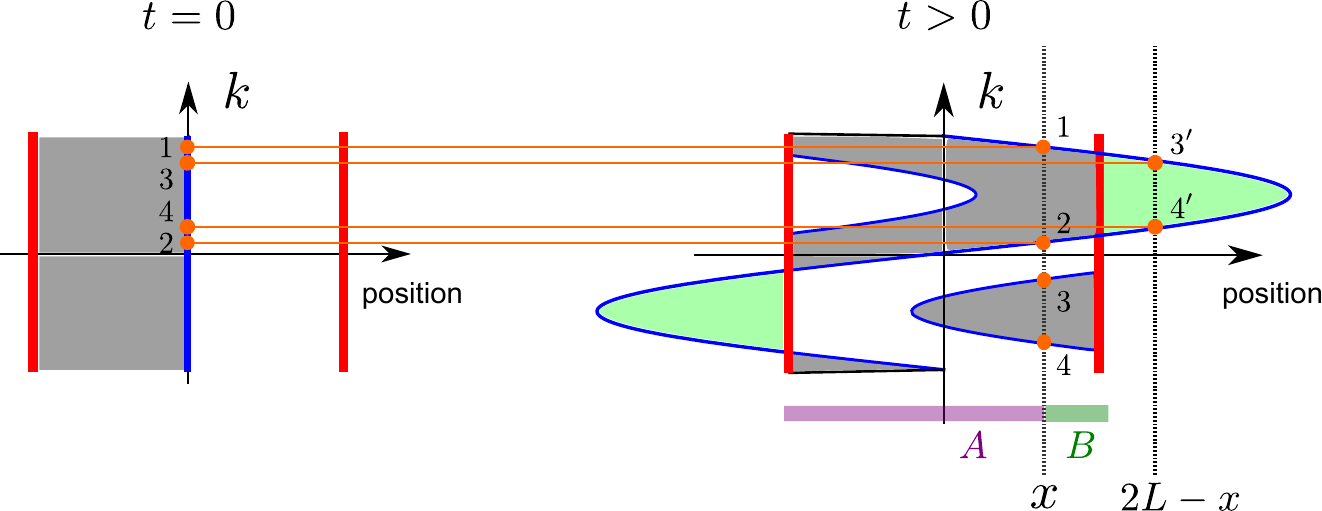}
\caption{Backward evolution of the Fermi points $\{k_{F,j}(x,t)\}_{j=1}^{4}$ at given entangling point $x$ and time $t>0$. Notice that 3 and 4 take an extra minus sign ($q_{j=3,4}=1$) due to the reflection occurred. }\label{fig:reflections}
\end{figure}

We now focus on the calculation of the R\'enyi entropy of a bi-partition $A\cup B$ with $A=[-L,x]$ and $B=(x,L]$ at fixed time $t$, as defined in Eq.~\eqref{eq:Renyi-def}.  In Sec.~\ref{sec:phase-space-dyn}, we showed that a cut in real space at position $x$ and time $t$ identifies a set of Fermi points $\{k_{F,j}(x,t)\}_{j=1}^{2Q}$ given by Eq.~\eqref{eq:Fermi-points-refl}. Due to momentum conservation, such Fermi points are transported backward in time up to the initial Fermi contour in Eq.~\eqref{eq:initial-Gamma} as \cite{Ruggiero2019,Scopa2021a,Scopa2021b,Ruggiero2022}
\be\label{eq:backward-evo}
\{k_{F,j}(x,t)\}_{j=1}^{2Q}\; \overset{\text{backward evo.}}{\longrightarrow}\; \{\theta_j\}_{j=1}^{2Q}=\{(-1)^{q_j} \ k_{F,j}(x,t)\}_{j=1}^{2Q},
\ee
 and the set $\{\theta_j\}_{j=1}^{2Q}$ can be used to identify the boundary points separating $A$ and $B$ along $\Gamma_t$, as discussed in the previous section. In Eq.~\eqref{eq:backward-evo},  $q_j$ is the number of reflections of the $j^{\rm th}$ particle at the walls given by
\be
q_j =\begin{cases} 0, \quad\text{if $|t\sin k_{F,j}|<L$};\\[4pt]
 q\in \mathbb{N}, \quad \text{if $(2q-1)L<|t\sin k_{F,j}|<(2q+1)L$}.\end{cases}
\ee
We illustrate the backward time evolution procedure in Fig.~\ref{fig:reflections}. \\

\begin{figure}[t]
\centering
\includegraphics[width=0.65\textwidth]{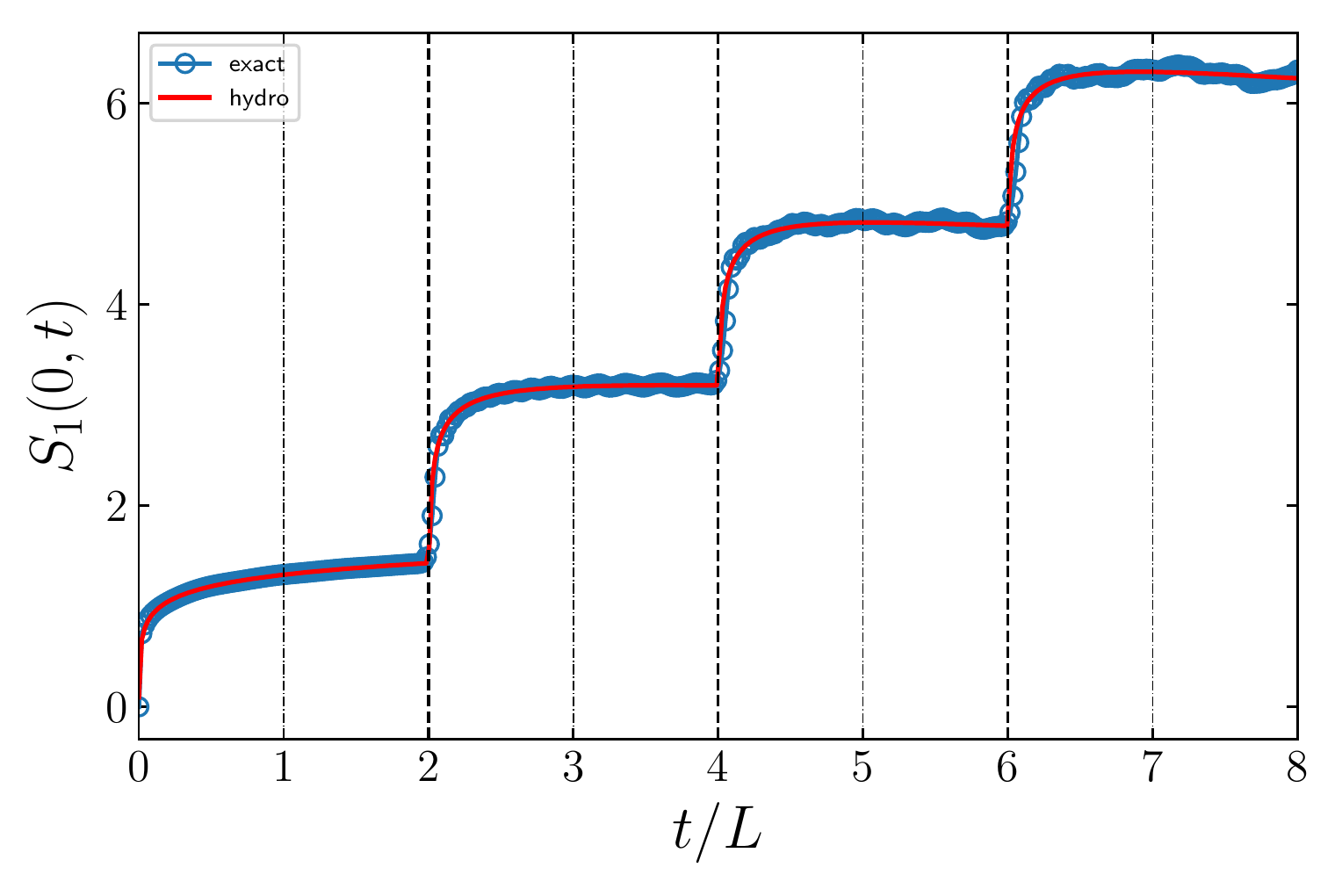}
\caption{Half-system entanglement entropy as function of time. The hydrodynamic prediction of Eq.~\eqref{eq:crazy} ({\it solid line}) is found in extremely good agreement with the numerical data obtained for a lattice of 300 sites ({\it symbols}). The dashed (resp.~dot-dashed) vertical axes mark the times $t=2\eta L$ (resp.~$t=(2\eta-1)L$), $\eta\in\mathbb{N}$, corresponding to the merging (resp.~reflection) of the left and right light cones.}\label{fig:S-half}
\end{figure}

Once determined the set of coordinates $\{\theta_j\}_{j=1}^{2Q}$ along $\Gamma_t$, we are ready to compute the R\'enyi entropy. In particular, as well-known from Refs.~\cite{Calabrese2004,Cardy2008,Calabrese2009}, the R\'enyi entropy \eqref{eq:Renyi-def} can be split as the sum of two contributions
\be\label{eq:S-ana}\begin{split}
&S_n(x,t)=\frac{1}{1-n}\log\left[\epsilon(x,t)^{\Delta_n}\braket{\hat{\cal T}_n(x,t)}\right]\\[4pt]
&\quad=\tilde{S}_n(x,t)- \frac{n+1}{12n}\log\epsilon(x,t) .
\end{split}\ee
The first term $\tilde{S}_n(x,t)$ is universal in the sense that it can be calculated within the effective field theory description of Eq.~\eqref{eq:H-LL} as the expectation value of a twist field $\hat{\cal T}_n$ inserted at the entangling point $x$. The twist field is a primary with scaling dimension $\Delta_n=\frac{1}{12}(n-1/n)$, see Refs.~\cite{Cardy2008,Calabrese2009} for a comprehensive discussion. 
Therefore, by decomposing $\hat{\cal T}_n$ into its chiral parts $\hat\tau^\pm_n$, with scaling dimension $d^\pm_n=\Delta_n/2$, and by performing the transformation of Eq.~\eqref{eq:backward-evo}, $\tilde{S}_n(x,t)$ can be determined from the $2Q$-point correlation function of $\hat\tau^\pm_n$ inserted at the points $\theta_j$

\be
\tilde{S}_n(x,t)=\frac{1}{1-n}\log\left[\prod_{j=1}^{2Q} \ \left|\frac{\dd \theta_j}{\dd x}\right|^{\nicefrac{\Delta_n}{2}}\left\langle\prod_{j=1}^Q \hat\tau_n^+(\theta_{2j-1})\hat\tau^-_n(\theta_{2j})\right\rangle\right].
\ee
The Jacobian factors can be explicitly written as (cf. Eq.\eqref{eq:Fermi-points-refl})
\be
\left|\frac{\dd \theta_j}{\dd x}\right|\equiv \left|\frac{\dd k_{F,j}(x,t)}{\dd x}\right|=\left[t\sqrt{1-R_j^2/t^2}\right]^{-1},
\ee
and $R_j$ associated to the Fermi points in Eq.~\eqref{eq:Fermi-points-refl}
\be 
R_\nu=\begin{cases} 
|x|+2L(\nu-1), \quad  \nu=1,\dots, N_+;\\[4pt]
|x|+2L(\nu-N_+-1), \quad  \nu=N_++ 1,\dots, 2N_+;\\[4pt]
|x|+2L(2N_+-\nu), \quad  \nu=2N_++1,\dots, 2N_+ + N_-;\\[4pt]
|x|-2L (2Q-\nu+1), \quad \nu=2 N_+ +N_- +1, \dots,2Q.\\[4pt]
\end{cases}
\ee
The $2Q$-point correlation function is given by \cite{Calabrese2004,Casini2005}
\be
\left\langle\prod_{j=1}^Q \hat\tau^+_n(u_j)\hat\tau^-_n(v_j)\right\rangle= \frac{\prod_{1\leq i<j\leq Q} g(u_j,u_i) g(v_j,v_i)}{\prod_{i,j=1}^{Q} g(v_j,u_i)}, \quad g(u,v)=\left|2\sin\frac{u-v}{2}\right|^{\Delta_n}\, .
\ee
 The second contribution comes instead from the short-distance regularization $\epsilon(x,t)$ of the field theory result and it is therefore non-universal. For the non-interacting spin chain, its value can be analytically derived with the Fisher-Hartwig conjecture (see e.g. Ref.~\cite{Jin2004,Calabrese2010,Ares2017}) and reads \cite{Scopa2021b,Ruggiero2022}
\be\label{eq:cutoff}
\log\epsilon(x,t)=\sum_{1\leq i < j\leq 2Q} (-1)^{i+j} \log\left|\sin\frac{k_{F,i}-k_{F,j}}{2}\right| +\frac{Q}{2} (\Upsilon_n+\log(2)/3)
\ee
with $\Upsilon_n$ a real number whose explicit expression can be found in Ref.~\cite{Jin2004}, for instance $\Upsilon_1\simeq 0.49502$.
\begin{figure}[t]
\centering
\includegraphics[width=0.315\textwidth]{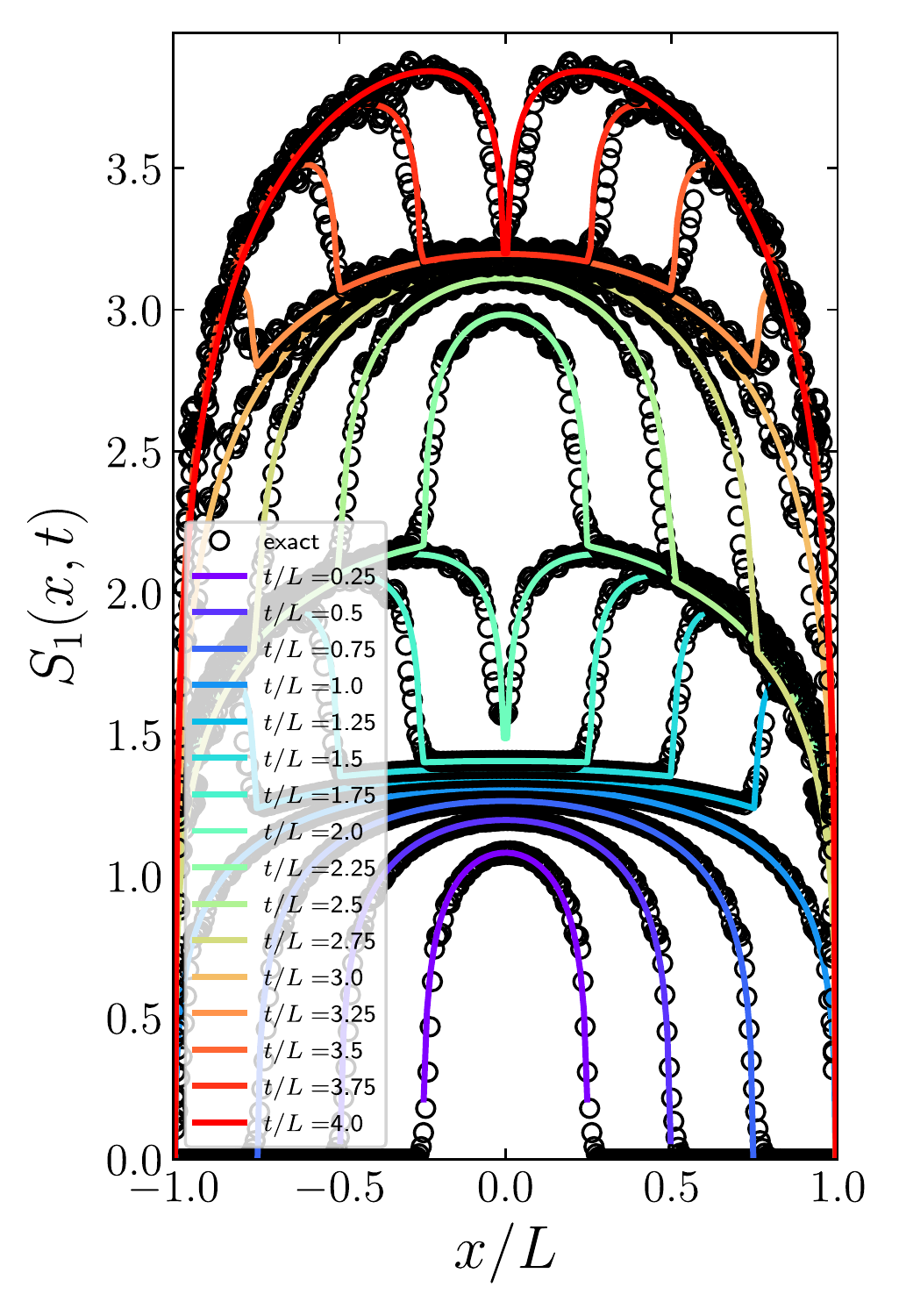}
\includegraphics[width=0.675\textwidth]{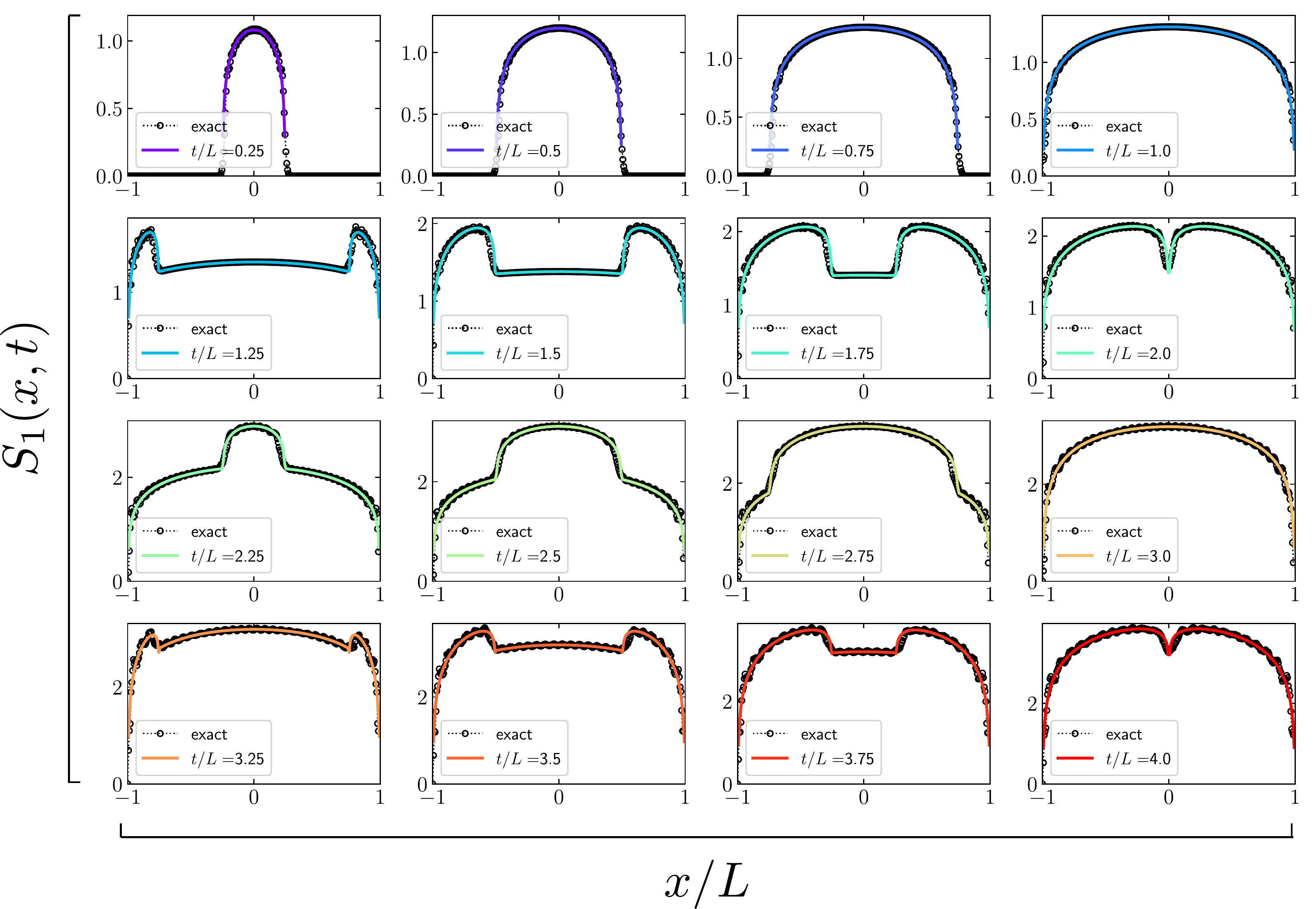}
\caption{Evolution of the entanglement entropy profiles during a QJE. Panels show different times $t$ increasing from the left to the right and from the top to the bottom, as written in each legend. The numerical data ({\it symbols}) obtained with exact lattice calculations on a lattice of $300$ sites is found in very good agreement with the hydrodynamic prediction in Eqs.~\eqref{eq:S-ana}-\eqref{eq:cutoff} ({\it solid line}) even after several reflections.}\label{fig:ent-refl}
\end{figure}
In Fig.~\ref{fig:ent-refl} we compare the hydrodynamic prediction for the entanglement entropy obtained combining Eqs.~\eqref{eq:S-ana}-\eqref{eq:cutoff} with exact lattice calculations, finding an extremely good agreement. Notice that the non-linear structure of the Fermi points \eqref{eq:Fermi-points-refl} does not allow for an easy manipulation of our results in order to get a simpler and closed expression of $S_n$. To give a hint of the complexity of this algebra, we report the expression for the entanglement entropy at half-system ($x=0$) and at given time $2qL\leq t<(2q+2)L$ after $q$ reflections have occurred,
\be\begin{split} \label{eq:crazy}
S_1(0,t)&=\frac{2q+1}{6}\log(2t)+\frac{2q+1}{2}\left(\Upsilon + \frac{\log 2}{3}\right)-\frac{q}{3}\log\left(\frac{t}{2 L}\right)-\frac{2}{3}\log\Gamma(q+1) \\[4pt]
 &\quad+\frac{1}{2}\log\left(\frac{t+2Lq}{t}\frac{\Gamma(t/(2 L) + q)}{\Gamma(t/(2 L) - q)}\right)-\frac{4}{3}\sum_{j=1}^q (-1)^j \log\tan\left(\frac{\varphi_j}{2}\right)
 \\[4pt]
 &\quad -\frac{4}{3}\sum_{1\leq j<k\leq q} (-1)^{j+k} \left[\log\tan\frac{\varphi_j+\varphi_k}{2} + \log\tan\frac{\varphi_k-\varphi_j}{2}\right],
\end{split}\ee
where $\Gamma(\cdot)$ is the Euler Gamma function and we defined the shorthand $\varphi_j\equiv \arcsin(2jL/t)$. In Fig.~\ref{fig:S-half}, we show the growth of entanglement at half-system obtained from Eq.~\eqref{eq:crazy} against numerical data. As one can see, the half-system entanglement displays some jumps at times $t=2\eta L$, $\eta\in \mathbb{N}$ ({\it dashed vertical lines}). These jumps are a consequence of the merging at $x=0$ of the two light cones that reflected against the walls in $x=\pm L$ at times $t=(2\eta-1)L$, compare with Fig.~\ref{fig:ent-refl} to visualize this process with the profile of entanglement. This pattern of reflections and merging suggests a Floquet picture of the long time dynamics in terms of the stroboscopic time $t\equiv\eta L$ that we discuss in Sec.~\ref{sec:long-time-S}.

\subsection{Joule expansion vs multiple domain wall setting}
\begin{figure}[t]
\centering
\includegraphics[width=\textwidth]{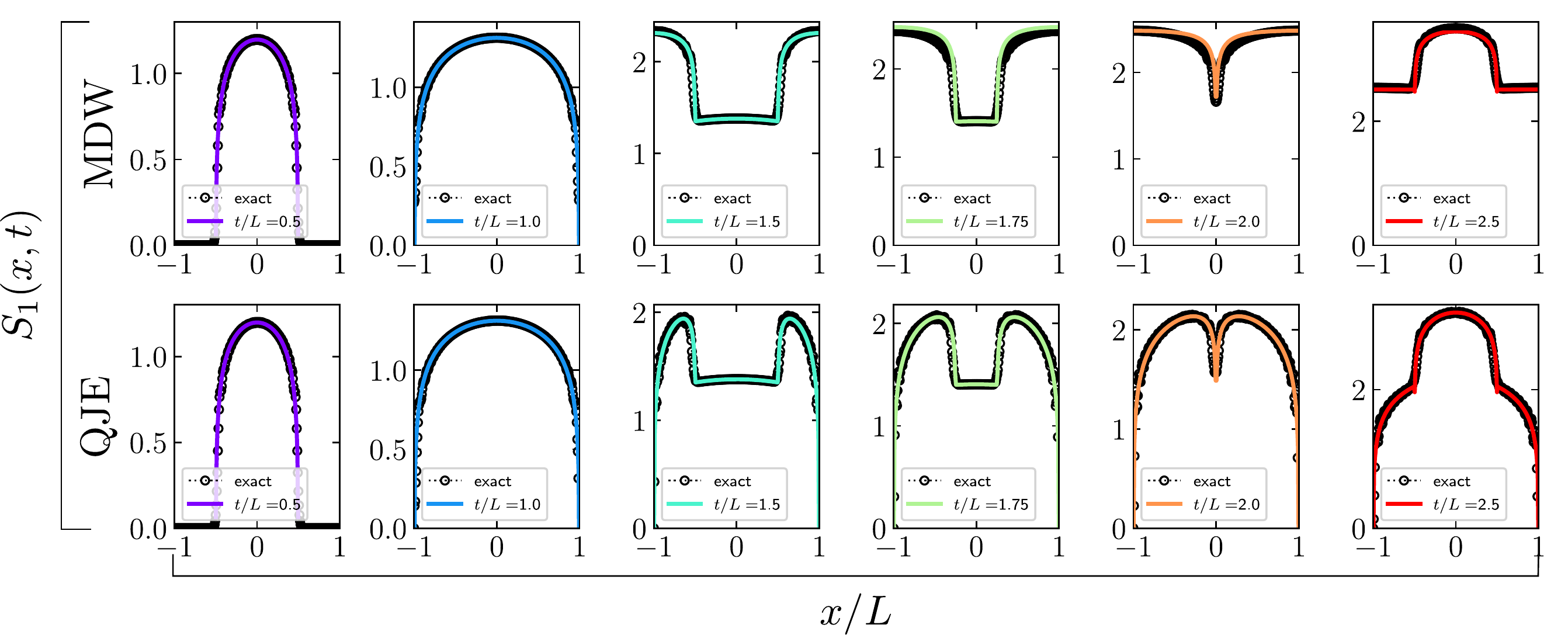}
\caption{Comparison of the entanglement dynamics in the MDW (top row) and in the QJE (bottom row) quench protocols as function of $x/L\in[-1,1]$ at different times increasing from the left to the right column. The two settings show the same entanglement profiles at times $t\leq L$ i.e., before any reflection, and significantly deviate at later times $t> L$. In both the settings, the analytical prediction obtained with quantum fluctuating hydrodynamics ({\it solid line}) is found in very good agreement with the numerical data ({\it symbols}). }\label{fig:comparison}
\end{figure}

It is interesting to notice that the entanglement shows a qualitatively different behavior in the quantum Joule expansion from that observed during the quench of a multiple domain wall configuration, even though the two settings are characterized by the same profile of conserved charges, see Sec.~\ref{sec:phase-space-dyn}. This feature is easily explained by noticing that, although the two settings share the same Fermi points $\{ k_{F,j}(x,t)\}_{i=1}^{2Q}$ in Eq.~\eqref{eq:Gamma-refl}, the latter come from different points on the initial Fermi contour, and so, they are characterized by very different quantum correlations in the two cases. More precisely, while the boundary points for the QJE are given in Eq.~\eqref{eq:backward-evo}, the dynamics of modes in the MDW is characterized by the absence of reflecting walls, therefore one can simply identify 
\be
\{\theta_j\}_{j=1}^{2Q} \equiv \{k_{F,j}(x,t)\}_{j=1}^{2Q}.
\ee 
Quite nicely, we notice that the factor $(-1)^{q_j}$ appearing in Eq.~\eqref{eq:backward-evo} is responsible for  the correct implementation of the reflecting walls in the QJE problem and, therefore, it alone distinguishes between the two quench problems, ensuring (if present) that the entanglement entropy drops at positions $x=\pm L$ because of the vanishing of the time-evolved wave function. In Fig.~\ref{fig:comparison} we show a comparison of the entanglement spreading in the two problems.

\section{Long time limit: reflections, relaxation and recurrence}\label{sec:long-time}
A natural question that might arise considering the Joule expansion protocol is about the fate of the quantum gas at very large times $t/L\gg1$, after it undergoes several reflections against the walls. Naively, one could imagine that the gas eventually relaxes towards a steady state where the system is in a homogeneous configuration with half-filling. Although this classical picture turns out to be somewhat correct for the conserved charges, understanding how this relaxation takes place and the asymptotic behavior of the entanglement is less intuitive. Moreover, due to the unitary character of the time evolution, revival effects are expected to manifest at large but finite times. For sake of clarity, it is useful to distinguish among three regimes occurring during the non-equilibrium dynamics at different time scales. Precisely, we find
\begin{itemize}
 \item[({\it i})] A reflection regime at $t/L \sim {\cal O}(1)$, i.e. at Euler scales, which is captured by the hydrodynamics developed in the previous sections. Here, we observe a Floquet picture of the large-time dynamics with periodicity $4L$.
\item[({\it ii})] An equilibration regime for $1\ll t/L \ll L$, where the system already went through several reflections and slowly relaxes towards a steady state configuration. In this regime, we observe a  breakdown of the hydrodynamic theory.  
\item[({\it iii})] A recurrence regime for $t\sim {\cal O}(L^2)$, over which the time-evolved wave function goes arbitrarily close to the initial configuration, as expected by standard ergodicity arguments. 
\end{itemize}
\begin{figure}[t]
\centering
({\it i}) \hspace{2.75cm} ({\it ii}) \hspace{2.75cm} ({\it iii})\\
\includegraphics[width=0.8\textwidth]{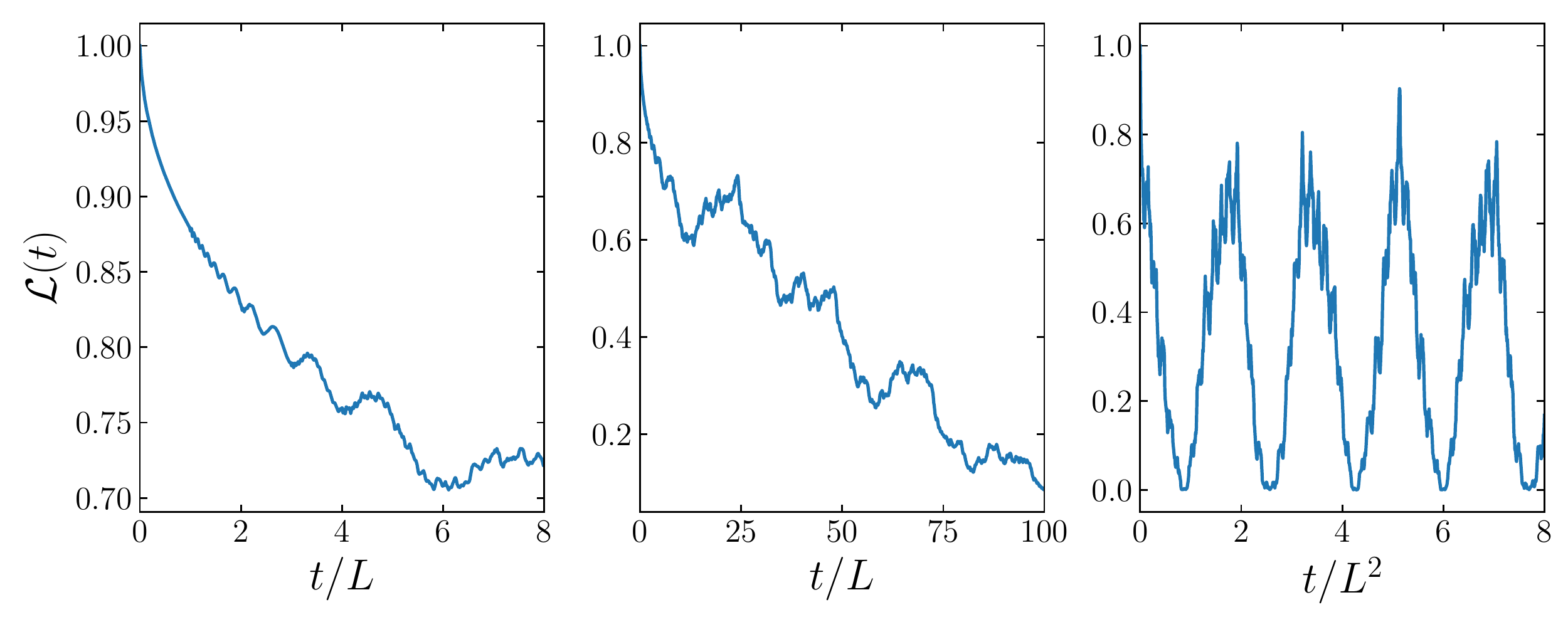}
\caption{Loschmidt echo ${\cal L}(t)$ in Eq.~\eqref{eq:echo} as function of time for different regimes: ({\it i}) reflection regime $t/L\sim {\cal O}(1)$; ({\it ii}) equilibration regime $1\ll t/L \ll L$ and ({\it iii}) recurrence regime $t/L\sim {\cal O}(L)$ from the left to the right panel. The curves are obtained numerically for a lattice of $300$ sites.}\label{fig:echo}
\end{figure}
In Fig.~\ref{fig:echo}, we plot the Loschmidt echo, defined as
\be\label{eq:echo}
{\cal L}(t)=\left|\braket{\Psi(t)|\Psi(0)}\right|^2,
\ee
at different time scales corresponding to the regimes ({\it i})-({\it iii}) discussed above. From this figure, it is easy to see that the revival time scale is $\sim {\cal O }(L^2)$ and, therefore, a stationary behavior can only be achieved in the scaling limit $t/L\to \infty$ such that $t/L^2\to 0$ \cite{Schutz1999,Collura2013b, Kaminishi2015,Rylands2019}.

\subsection{Long time limit of the density}
We first focus on the particle density starting from the result in Eq.~\eqref{eq:dens-copy}, which is a particularly pleasant form for the study of the long time limit. Indeed, one can asymptotically replace the series appearing in Eq.~\eqref{eq:dens-copy} using the Euler-MacLaurin formula as
\be\label{eq:dens-fili}
I(\pm x,t)\equiv \sum_{p=1}^{N_\mp} (-1)^{p}\rho(x_p\mp x,t)\approx \int_{1}^{N_\mp} \dd p \ f(p) + \frac{f(N_\mp)+f(1)}{2} + \text{subleading terms}
\ee
where $f(p)=\cos(\pi p)\bar\rho(x_p\mp x,t)$. It is then easy to show that $\lim_{t\to\infty} I(x,t) = 1/4$ and, from Eqs.~\eqref{eq:dens-0refl}-\eqref{eq:dens-fili}, to conclude that
\be\label{eq:dens-asy}
\lim_{t\to \infty}\rho(x,t)=\lim_{t\to \infty}\bar\rho(x,t) = 1/2.
\ee

\begin{figure}[t]
\centering
\includegraphics[width=0.8\textwidth]{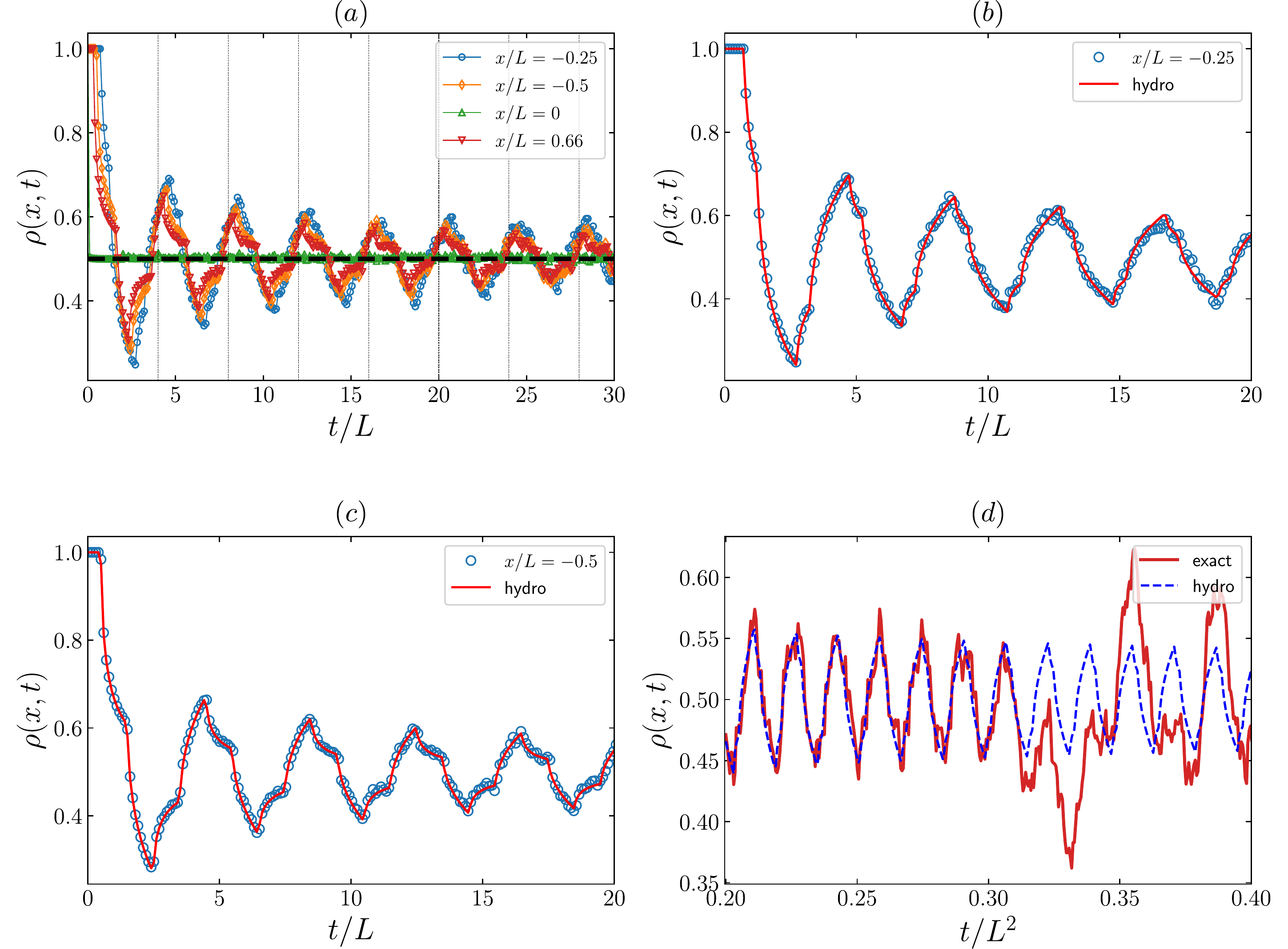}\\
\caption{Panel $(a)$: Particle density dynamics at fixed position $x$ as function of time, obtained with exact numerical calculations on a lattice of $500$ sites. The horizontal dashed line marks the value $\rho=1/2$ while vertical axes are drawn at each period $4L$. Panel $(b)$-$(c)$: Same plot where the numerical data ({\it symbols}) are compared with the hydrodynamic prediction in Eq.~\eqref{eq:dens-refl} ({\it solid line}) for two different values of $x$. We can see that by considering the subleading contributions in Eq.~\eqref{eq:dens-fili}, hydrodynamics captures the oscillations of density, going beyond the classical result. Panel $(d)$ shows the breakdown of hydrodynamics at times $t/L\sim 0.3 L$ for $x/L=-0.25$ and for the choice of parameters considered. }\label{fig:dens-asy}
\end{figure}

The result in Eq.~\eqref{eq:dens-asy} agrees with the expectation for the asymptotic density in a classical Joule experiment. However, in Fig.~\ref{fig:dens-asy}$(a)$, we show that the underlying unitary evolution manifests in persistent oscillations  around the classical result $\rho=1/2$ with period $4L$ and decreasing amplitude, see also Ref.~\cite{Collura2013,Collura2013b}. These oscillations are well captured by the subleading terms appearing in Eq.~\eqref{eq:dens-fili} that we previously neglected to recover the classical result in \eqref{eq:dens-asy}, see Fig.~\ref{fig:dens-asy} -- panels $(b)$ and $(c)$.  We numerically observe the same qualitative behavior up to the equilibration regime $1\ll t/L \ll L$, after which hydrodynamics deviates from the numerical data, see Fig.~\ref{fig:dens-asy}~$(d)$.\\
\begin{figure}[t]
\centering
({\it i}) \hspace{3cm} ({\it ii}) \hspace{3cm} ({\it iii$_1$}) \hspace{3cm} ({\it iii$_2$}) \\
\includegraphics[width=\textwidth]{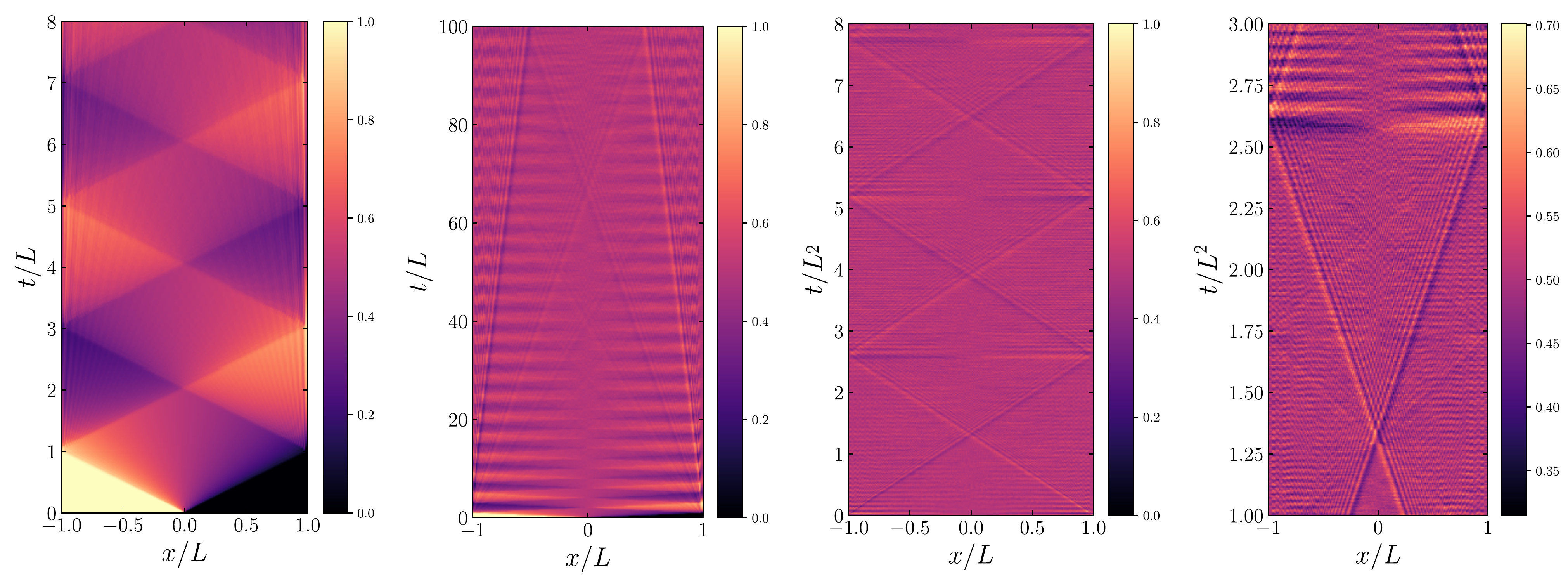}
\caption{Colormap plot of the particle density $\rho(x,t)$ in the $x$-$t$ plane at different time scales, increasing from panel ({\it i}) to panel ({\it iii}). The data is obtained with exact lattice calculations on a system of $300$ sites.}\label{fig:dens-revival}
\end{figure}
We provide a thorough numerical analysis of the particle density profile in Fig.~\ref{fig:dens-revival}, showing $\rho(x,t)$ as a colormap plot in the $x$-$t$ plane at different time scales. In panel ({\it i}), corresponding to the reflection regime, we clearly see the pattern of density oscillations with $4L$ periodicity. The latter gradually smooth out on larger time scales corresponding to the equilibration regime where $\rho \approx 1/2$ homogeneously along the chain, as shown in panel ({\it ii}). Finally, new patterns in the density profile arise at $t\sim {\cal O}(L^2)$ and correspond to those times where ${\cal L}(t) \sim 1$, i.e. to recurrences, or to times where the wave function is orthogonal to the initial state and ${\cal L}(t)\sim 0$, compare Fig.~\ref{fig:dens-revival} ({\it iii$_1$}) with Fig.~\ref{fig:echo}. Notice that the structure of the patterns in the recurrence regime is very rich and some of the local minima and maxima of ${\cal L}(t)$ correspond to the sub-patterns shown in Fig.~\ref{fig:dens-revival} ({\it iii$_2$}).

\subsection{Long time limit of the entanglement}\label{sec:long-time-S}

\begin{figure}[t]
\centering
\includegraphics[width=0.535\textwidth]{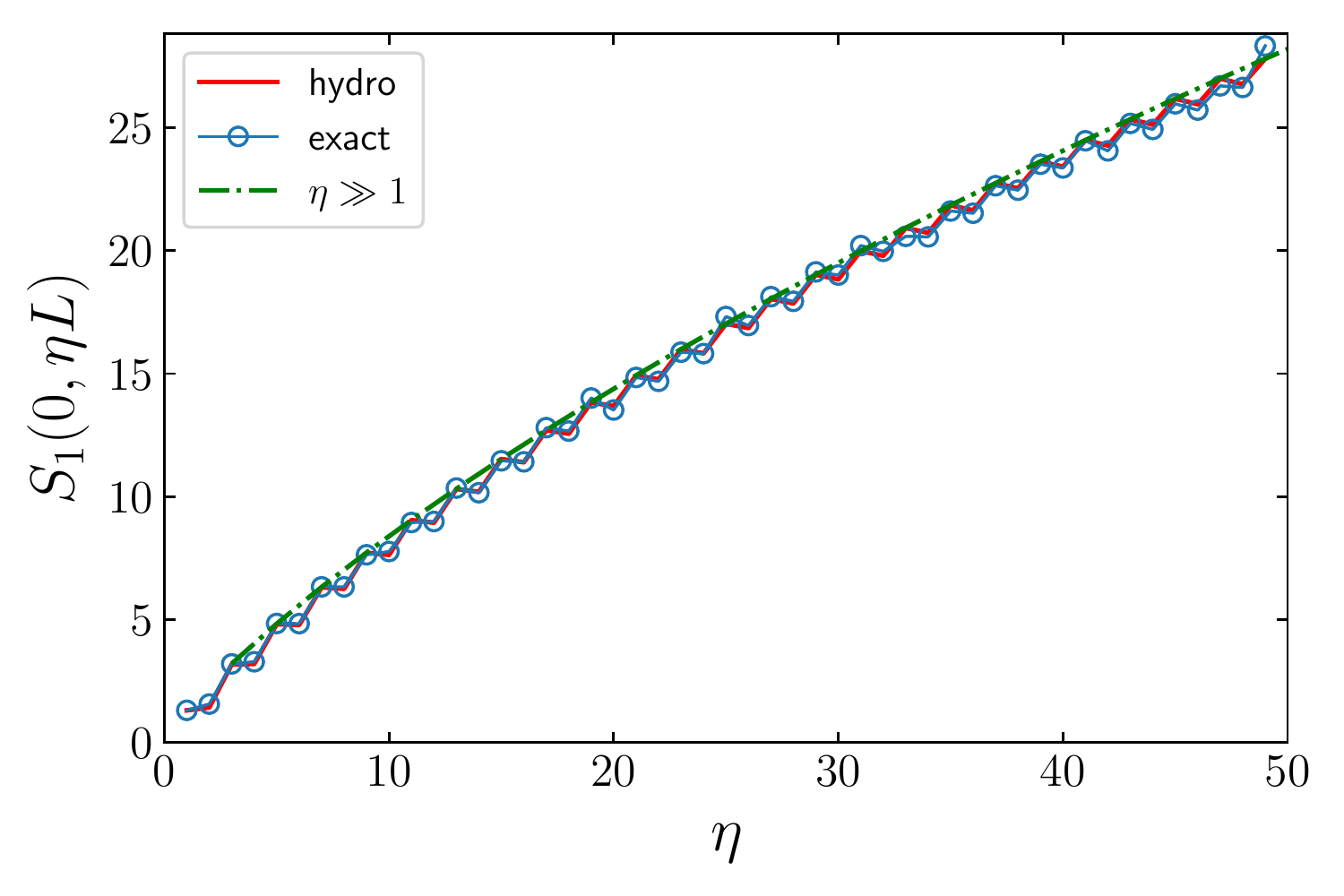}
\caption{Stroboscopic growth of the half-system entanglement $S_1(0,\eta L)$ with $\eta\in\mathbb{N}$. The figure shows the exact numerical data ({\it symbols}) for a lattice of $300$ sites, the hydrodynamic prediction of Eq.~\eqref{eq:crazy} ({\it solid line}) and its asymptotic behavior given in Eq.~\eqref{eq:S-ansatz} ({\it dot-dashed line}), which are in extremely good agreement even after many periods.}\label{fig:long-EE1}
\end{figure}

\begin{figure}[t]
\centering
$(a)$ \hspace{4cm} $(b)$ \\
\includegraphics[width=0.35\textwidth]{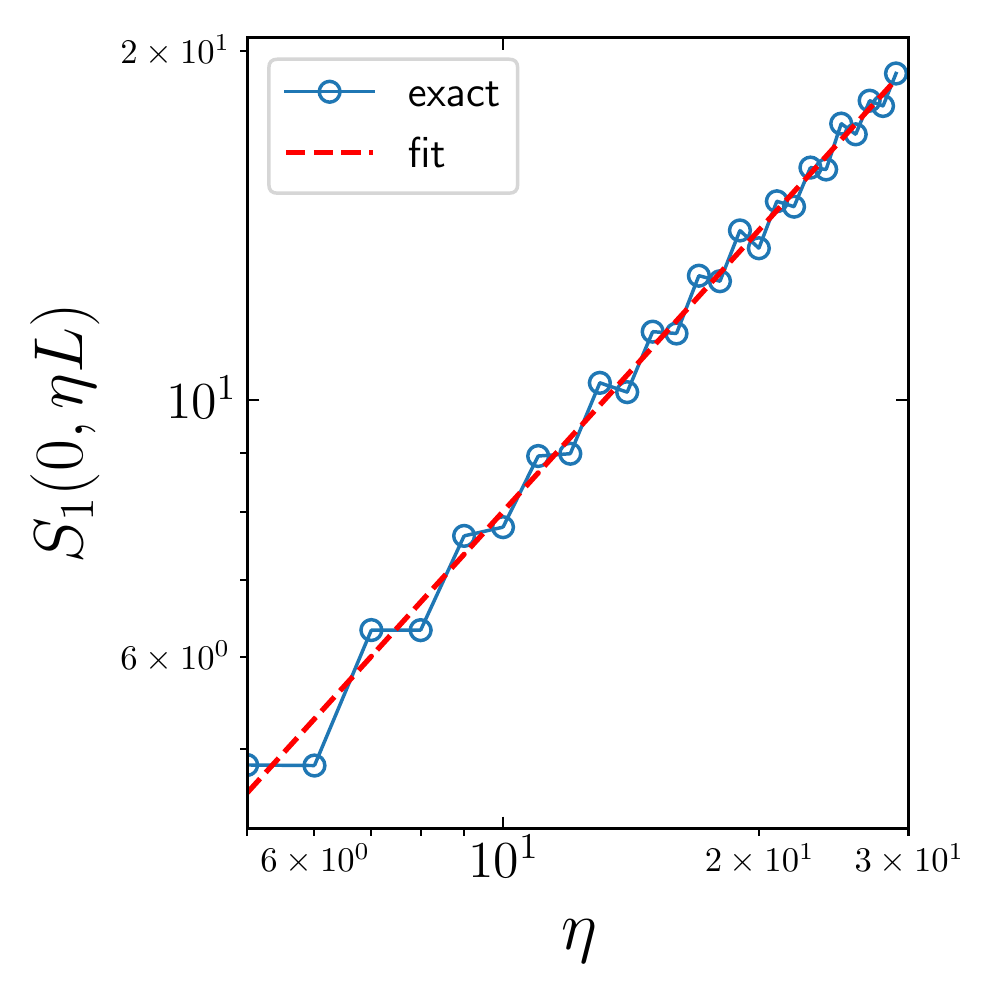}\includegraphics[width=0.53\textwidth]{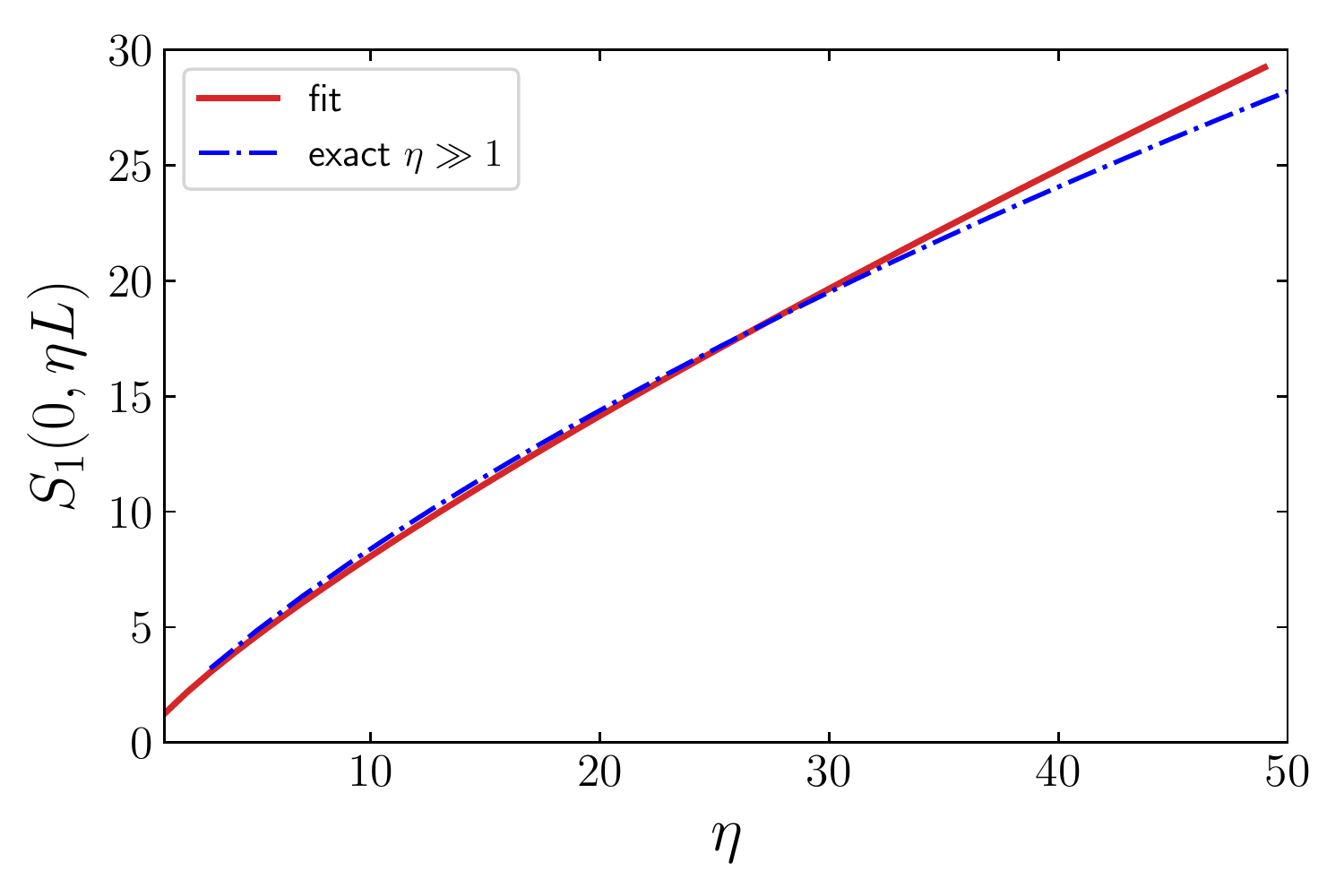}
\caption{Panel $(a)$ --- Stroboscopic half-system entanglement in a log-log plot ({\it symbols}) and the fitting function $S_1(0,\eta L)\sim C \eta^\alpha$ ({\it dashed line}) with $\alpha\simeq 0.81$ and $C\simeq 1.25$ for our choice of parameters. Panel $(b)$ --- The fitting function is compared to the asymptotic result in Eq.~\eqref{eq:S-ansatz} of the stroboscopic entanglement for $\eta\gg 1$. }\label{fig:long-EE}
\end{figure}

We finally discuss the long time behavior of the entanglement entropy. As already commented in Sec.~\ref{sec:entanglement-calc}, from Fig.~\ref{fig:ent-refl} and \ref{fig:S-half} one can easily see that the entanglement dynamics during the QJE  is made of two alternating stages: {\bf 1.}~the spreading of the domain wall from $x=0$ to $x=\pm L$, where it reflects against the walls; {\bf 2.}~the dynamics of the two reflected (half) domain walls from $x=\pm L$ to $x=0$ and their subsequent merging. Focusing on the asymptotic dynamics at half system $x=0$, we notice that these reflection (merging) processes occur at odd (even) multiples of the rescaled time $\eta=t/L$, hence suggesting a Floquet picture of the long-time dynamics in terms of the stroboscopic time $\eta$. We show this analysis in Fig.~\ref{fig:long-EE1}, where the exact lattice numerics matches the hydrodynamic prediction with excellent agreement in the reflection regime $\eta\sim{\cal O}(1)$. 
Moreover, a careful analysis of Eq.~\eqref{eq:crazy} reveals that the half-system entanglement grows for $\eta\gg 1$ as  
\be\label{eq:S-ansatz}
S_1(0,\eta L) \overset{\eta\gg 1}{\sim} 2\alpha(L) \eta -\frac{2}{3}\eta \log\eta -\frac{1}{6}\log\eta   
\ee
up to an additive constant, with
\be
\alpha(L)=\frac{1}{3}\log\left(\frac{4L}{\pi^2}\right)+\Upsilon+\frac{1}{3}.
\ee
In Fig.~\ref{fig:long-EE1}, we observe that Eq.~\eqref{eq:S-ansatz} captures extremely well the behavior of entanglement even at relatively small values of $\eta$. In the absence of the analytical result in Eq.~\eqref{eq:crazy}, one might conjecture  a power-law behavior $S_1\sim C(t/L)^a$ for the asymptotic growth (see e.g. Ref.~\cite{Alba2014}), which we find indeed numerically very close to the correct result in Eq.~\eqref{eq:S-ansatz}, see Fig.~\ref{fig:long-EE}.  

\begin{figure}[t]
\centering
$(a)$ \hspace{4cm} $(b)$\\
\includegraphics[width=0.85\textwidth]{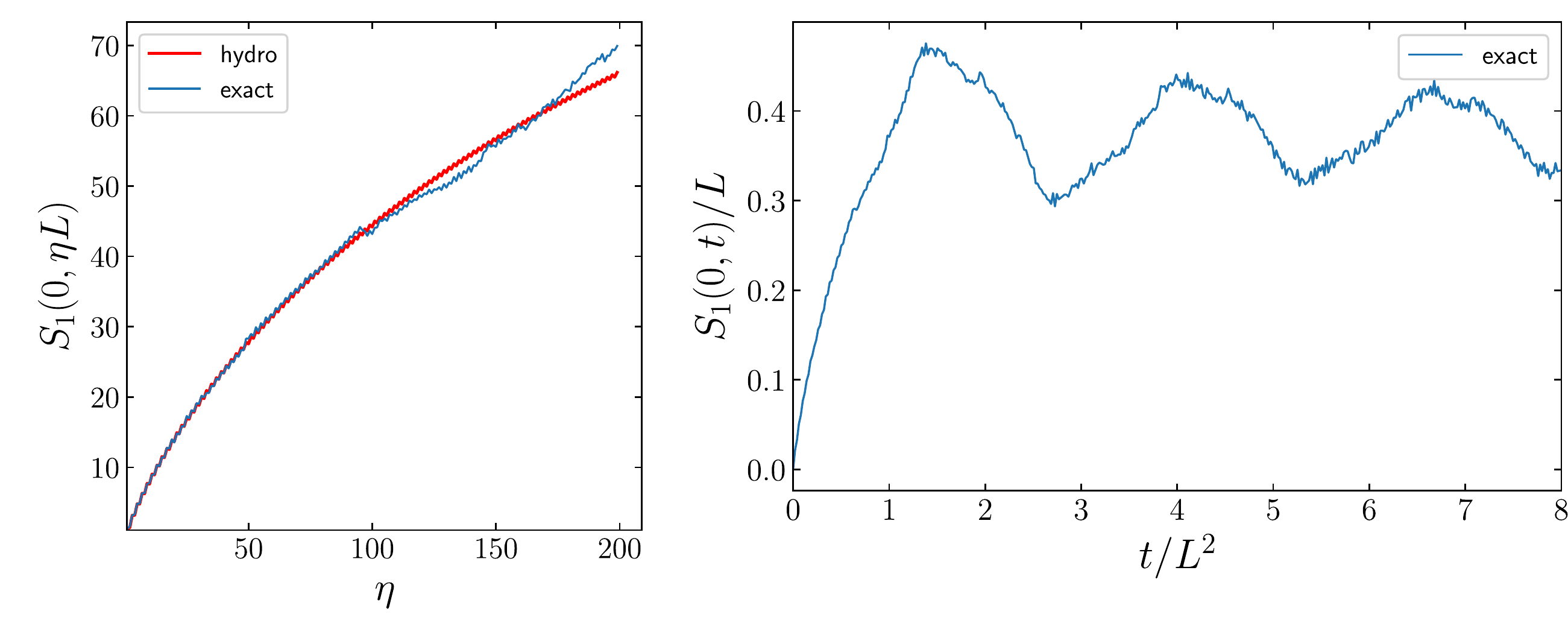}
\caption{Asymptotic behavior of the half-system entanglement beyond the regime of hydrodynamics. Panel $(a)$ -- Stroboscopic half-system entanglement obtained with exact numerics on a lattice of $300$ sites ({\it pale blue line}) significantly deviates from hydrodynamics ({\it red line}) at times $t/L\sim 100$ with our choice of parameters. Panel $(b)$ -- Half-system entanglement entropy in the recurrence regime.}\label{fig:EE-recurrence}
\end{figure}

Moving to larger time scales $1\ll \eta\ll L$ we still observe a growth of the entanglement associated with the relaxation dynamics in place but we also witness to a significant deviation of $S_1(0,t)$ from the hydrodynamic prediction, see Fig.~\ref{fig:EE-recurrence}$(a)$. Concluding, in Fig.~\ref{fig:EE-recurrence}$(b)$ we investigate the regime $t\sim {\cal O}(L^2)$ where we register oscillations of the entanglement that correspond to the local minima and maxima of ${\cal L}(t)$ in Fig.~\ref{fig:echo} and, consequently, to the pattern in Fig.~\ref{fig:dens-revival} ({\it iii$_1$}) observed for the particle density, see also Ref.~\cite{Alba2014}. 

\section{Summary and conclusion}\label{sec:conclusion}
We investigated the non-equilibrium dynamics of a one-dimensional gas of spinless fermionic particles with nearest-neighbor hopping Hamiltonian, initially prepared in a box of size $L$ with saturated density $\rho=1$ and subsequently let to expand in a doubled sized region by suddenly moving one of the box's edges. This simple quench protocol might be seen as a quantum version of the Joule expansion, thus it was dubbed as quantum Joule expansion in our analysis. For this problem, we employed tools stemming from phase-space hydrodynamics to reconstruct the profile of conserved charges at Euler space-time scales and subsequently, following the recent literature on quantum fluctuating hydrodynamics \cite{Ruggiero2019,Ruggiero2020,Scopa2021a,Scopa2021b,Ruggiero2022,Rottoli2022}, we re-quantized the hydrodynamic background to obtain an analytical prediction for the R\'enyi entropies during the quench dynamics, which we tested against numerics for the von Neumann entanglement entropy. Remarkably, we found that the prediction of quantum hydrodynamics is able to reproduce the numerical data obtained for the lattice model with great accuracy even after several reflections of the quantum gas.\\
 In the last part of this work, we studied the long time behavior of the quantum gas, building on previous results appeared in literature, see e.g. Ref.~\cite{Schutz1999,Platini2007,Collura2013,Collura2013b}. We observed three time scales that characterize the long time physics and we provided a careful analysis of each. First, we found a reflection regime $t/L\sim{\cal O}(1)$ where the quench dynamics is still well captured by our hydrodynamic theory. Here, we discussed the persistent oscillations of the particle density around the classical value $\rho=1/2$ and we calculated the half-system entanglement growth for $t\gg L$ to be the non-trivial function $S_1 \sim 2\alpha(L) t/L -\frac{2t}{3L} \log (t/L) -\frac{1}{6}\log(t/L)$, extracted from the asymptotic expansion of our analytical result. At larger scales, we found an equilibration regime $1\ll t/L\ll L$ where hydrodynamics breaks down and the system relaxes to a stationary state and, finally, a recurrence regime for $t\sim{\cal O}(L^2)$ characterized by revivals of the many-body wave function. \\
 \indent
We envisage at least two follow-ups of this work. One future direction could be the study of the entanglement spreading in systems with boundaries for interacting integrable models (see e.g.~\cite{Grochowski2021})
or for more generic choice of confinement. On the other hand, it would be interesting to better explore the long time regime where hydrodynamics breaks down and the system eventually relaxes to a non-equilibrium steady state, see e.g. Ref.~\cite{Joule-exp1,Joule-exp2}. This path would probably require to take into account higher order corrections to the Euler hydrodynamics not considered in this work, see \cite{Fagotti2017,Fagotti2020}.

\vspace{0.2cm}
{\it Acknowledgments ---} FA and SS acknowledge support from ERC under Consolidator grant number 771536
(NEMO). The authors are thankful to Dragi Karevski, Pasquale Calabrese and J\'er\^ome Dubail for useful discussions and insights on the manuscript. FA acknowledges Nina Javerzat for helpful comments on Eq.~\eqref{eq:S-ansatz}. SS acknowledges LPCT (Nancy) for the warm hospitality during the early stage of this project.
\appendix

\section{Calculation of the Fermi contour}\label{app:Fermi-contour}
\begin{figure}[t]
\centering
$(a)$ \hspace{3.5cm} $(b)$ \hspace{3.5cm} $(c)$\\
\includegraphics[width=.3675\textwidth]{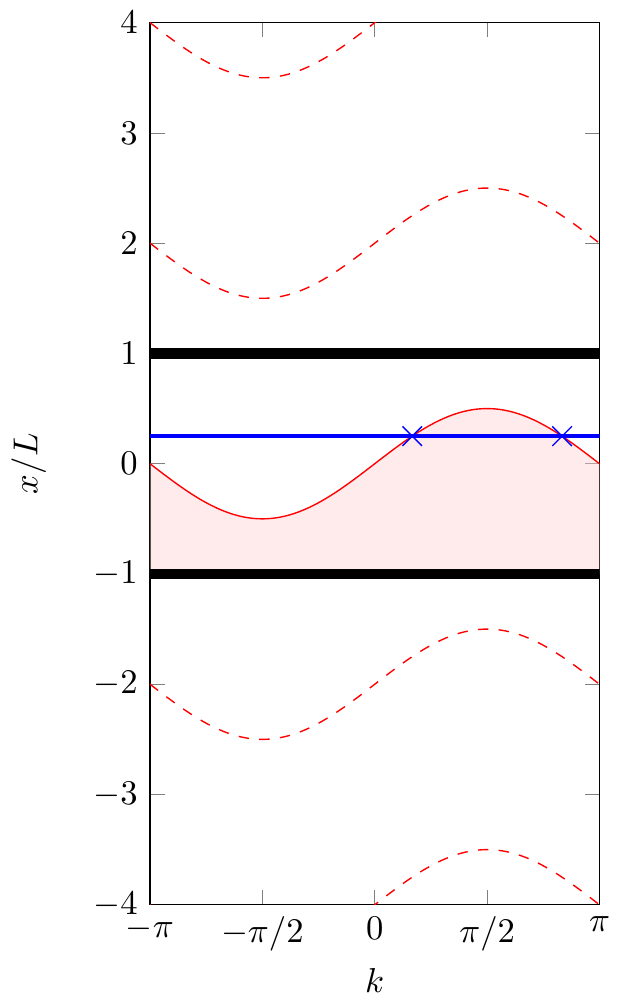}
\includegraphics[width=.3\textwidth]{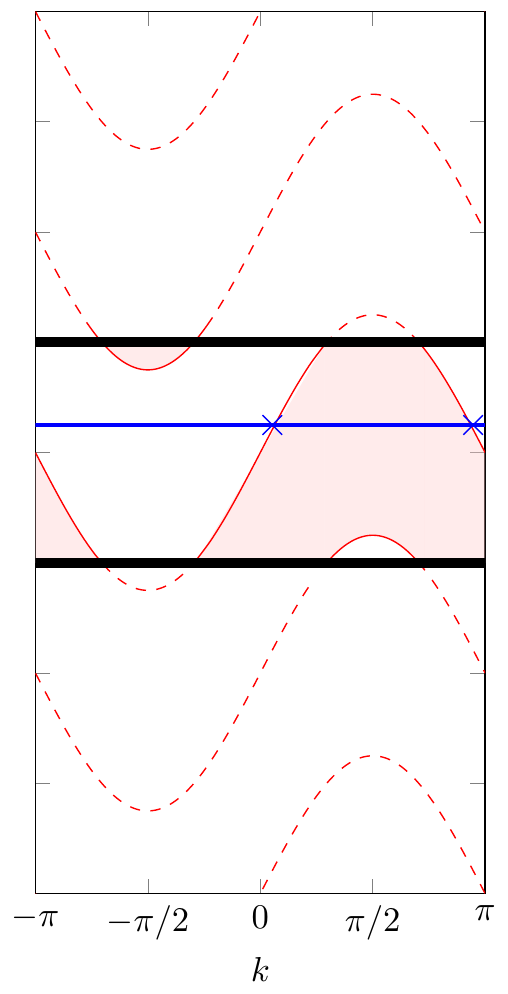}
\includegraphics[width=.3\textwidth]{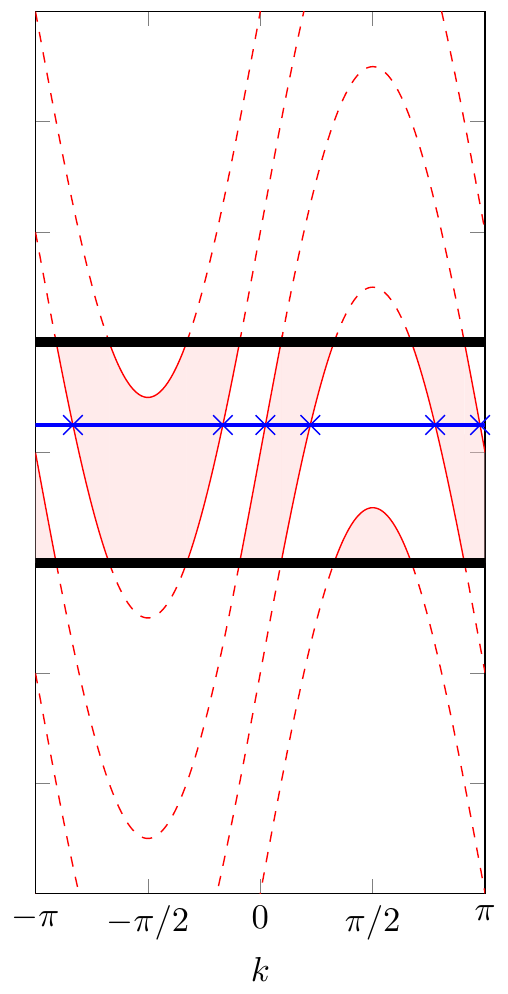}
\caption{ Illustration of the time evolution of the Fermi contour --
Snapshots of the time evolution of the Fermi contour after the QJE.
Panel $(a)$ -- $0<t<L$, no reflection has taken place yet 
and the quantum gas is freely expanding. Panel $(b)$ --
$L<t<2L$, the first reflection has taken place and the Fermi sea is
split. Panel $(c)$ -- $3L<t<5L$, the fastest modes 
have undergone two reflections.
The blue line for $x=0.25L$ illustrates that the 
solution~(\ref{eq:app-kf}) correctly reproduces the edges of the Fermi surface.
}
\label{fig:images}
\end{figure}

After the sudden expansion of the box confinement, the propagating modes on the initial Fermi contour $x=0$, $k\in [-\pi,\pi]$ in Eq.~(\ref{eq:initial-macrostate}) move freely with velocity $v(k)=\sin k$ up to the position of the two walls at  $x=\pm L$, where their trajectory is elastically reflected, $k\to -k$. As we have illustrated in Fig.~\ref{fig:Gamma-reflections}, each of these reflections further splits the Fermi contour. Focusing on a position $x\in[-L,L]$ inside the box, we may employ the method of images and consider the free propagation of a system with infinitely many virtual copies of the Fermi contour rather than tracking the true reflection dynamics in the presence of the walls, see Fig.~\ref{fig:multipleDW}. In this picture, while the initial modes are allowed to leave the system, virtual modes will enter the system as to mimic the reflection dynamics, see Fig.~\ref{fig:images} for an illustration. The family of images is parametrized by the equation of motion of the modes generated on each of the junctions at $x_p=2Lp$ that arrived at position $x$ at a certain time $t$
\begin{align}\label{eq:app-fermi1}
x =x_p + t \sin(k)
\end{align}
where $p$ is an integer number. 
We can find out the number of virtual sources  that are activated following the trajectory of the light cone $x=2pL \pm t$, obtained from Eq.~\eqref{eq:app-fermi1} with $k=\pm\pi/2$. Considering $x>0$, we find that there are $N_-$ Fermi points in the interval 
$[-\pi, 0]$ with 
%
\begin{align}
	N_- = \left\lfloor\frac{t+x}{2L}\right\rfloor
\end{align}
and $N_+$ Fermi points in the interval $[0,\pi]$ with 
\begin{align}
    N_+ = \left\lceil\frac{t-x}{2L}\right\rceil.
\end{align}
Now that we know the valid range for $p$, we may solve for the Fermi 
points using Eq.~(\ref{eq:app-fermi1}) but straightforwardly solving will 
yield the Fermi points ordered by reflections and thus we need to rearrange
them such that they are ordered by magnitude.
We readily find
\begin{align}\label{eq:app-kf}
	k_{F,j}(x,t) = \begin{dcases}
	        	\arcsin\left( \frac{x + 2Lj}{t} \right), \ &j = -N_-,...,N_+-1\\
	        	\pi -\arcsin\left( \frac{x + 2L(2N_+ -1 - j)}{t} \right), \ & j = N_+,...,2N_+-1\\
	        	-\pi -\arcsin\left( \frac{x - 2L(j + 2N_- +1)}{t} \right), \ & j = -2N_-,...,-N_- -1
	        \end{dcases},
\end{align}
which are related to Eq.~\eqref{eq:Fermi-points-refl} of Sec.~\ref{sec:phase-space-dyn} by a simple re-definition of the index $j$.

\section{Exact lattice calculations}\label{app:numerics}
In this appendix, we briefly report the strategy of the exact numerical lattice calculations that we performed to generate the data in the main text. First, we write the free Fermi Hamiltonian in Eq.~\eqref{eq:model} as 
\be
\Ha= \sum_{i,j=1}^{N-1} \hat{c}^\dagger_j \ \mathbb{H}^{(0)}_{ij} \ \hat{c}_i ,
\ee
with $N$ the number of lattice sites and Hamiltonian matrix
\be\label{eq:H-matrix}
 \mathbb{H}^{(0)}_{ij}=-\frac{1}{2}\left(\delta_{i,j+1} +\delta_{i+1,j}\right) +V(j,t<0) \delta_{i,j}.
\ee
Next, we diagonalize the matrix in Eq.~\eqref{eq:H-matrix} as 
\be
\mathbb{H}^{(0)}=\bm{w}^\dagger \ \text{diag}(\vec{E}^{(0)}) \   \bm{w}, \quad \bm{w}=\bV \vec{w}_1 \big| \vec{w}_2 \big| \dots \big| \vec{w}_N\eV; \quad \vec{E}^{(0)}=(E_1^{(0)},\dots,E^{(0)}_N)
\ee
with $\vec{w}_n$ (resp.~$E^{(0)}_n$) the single-particle eigenvector (resp.~energy), and we construct the ground state of the model $\ket{\Psi(0)}$ by adding to the vacuum the negative-energy modes of the spectrum. The two-point correlation matrix is then obtained as
\be\label{eq:2pt-initial}
\mathbb{G}(0)=\left[ \braket{\Psi(0)| \hat{c}^\dagger_i \hat{c}_j|\Psi(0)}\right]_{i,j=1}^N = \bm{w} \ \Pi \ \bm{w}^\dagger
\ee
where $\Pi$ is a ground-state projector, defined as
\be
\Pi=\text{diag}(p_1,\dots, p_N), \quad p_j=\Theta(-E_j^{(0)})
\ee
and $\Theta$ is the Heaviside step function. At times $t>0$, we study the quench dynamics generated by the (post-quench) Hamiltonian matrix 
\be
\mathbb{H}_{i,j}=-\frac{1}{2}\left(\delta_{i,j+1} +\delta_{i+1,j}\right) +V(j,t>0) \delta_{i,j},
\ee
which is diagonalized as
\be
\mathbb{H}=\bm{v}^\dagger \ \text{diag}(\vec{E}) \   \bm{v}
\ee
with energies $E_n$ and eigenstates $\vec{v}_n$. It is then easy to exactly evolve the two-point function in Eq.~\eqref{eq:2pt-initial} in the basis of post-quench eigenstates, obtaining
\be
\mathbb{G}(t)=\bm{v} \ U(t) \ \bm{v}^\dagger \ \mathbb{G}(0)\ \bm{v} \ U^\dagger(t) \ \bm{v}^\dagger, \qquad U(t)=\text{diag}\left(\exp(-\I t \vec{E})\right).
\ee
From the knowledge of the two-point correlation matrix, other quantities can be derived exploiting Wick's theorem. In particular, the particle density is read from the diagonal elements of the correlation matrix
\be
\rho(j,t)=\delta_{i,j} \mathbb{G}_{i,j}(t).
\ee
and the entanglement entropy of a bi-partition $[1,\ell]\cup[\ell+1,N]$ is obtained from the eigenvalues $\zeta_i$ ($i=1,\dots,\ell$) of the minor $\mathbb{G}^{(\ell)}(t)=[\mathbb{G}_{i,j}(t)]_{i,j=1}^\ell$ as \cite{p-12,Peschel1999a,Chung2001,Peschel2003,Peschel2004,Peschel2009}
\be
S_1(\ell,t)=-\sum_{i=1}^\ell \left(\zeta_i\log\zeta_i + (1-\zeta_i)\log(1-\zeta_i)\right).
\ee

\section*{References}

\input{bibliography}
\end{document}

%% file: new_commands.tex
\newcommand*{\ddt}[1]{%
  \accentset{\mbox{\bfseries .\hspace{-0.25ex}.}}{#1}} 

\newcommand{\be}{\begin{equation}}
\newcommand{\ee}{\end{equation}}

\newcommand{\beA}{\begin{align}}
\newcommand{\eeA}{\end{align}}

\newcommand{\bV}{\begin{pmatrix}}
\newcommand{\eV}{\end{pmatrix}}

\newcommand{\dd}{\mathrm{d}}
\newcommand{\de}{\partial}

\newcommand{\Ha}{\hat{H}} 

